\documentclass[aps,preprint,floats,epsf,epsfig,nofootinbib,letter]{revtex4}

\usepackage{graphicx}
\usepackage{dcolumn}
\usepackage{bm}
\usepackage{subfigure}

\def\be{\begin{eqnarray}}
\def\en{\end{eqnarray}}
\def\non{\nonumber}

\def\Zslash{\!\not{\! Z}}
\def\eslash{\!\not{\! \epsilon}}

\begin{document}

\renewcommand{\baselinestretch}{1.10}

\font\el=cmbx10 scaled \magstep2{\obeylines\hfill Oct., 2012}

\vskip 1.5 cm

\centerline{\Large\bf   Implications of ${\cal B}(\mu\to e\gamma)$ and $\Delta a_\mu$}
\centerline{\Large\bf   on Muonic Lepton Flavor Violating Processes}

\bigskip
\centerline{\bf Chun-Khiang Chua}
\medskip
\centerline{Department of Physics, Chung Yuan Christian University}
\centerline{Chung-Li, Taiwan 320, Republic of China}
\medskip

\centerline{\bf Abstract}
We study the implications of the experimental results on the $\mu\to e\gamma$ decay rate and the muon anomalous magnetic moment, 
on muonic lepton flavor violating processes, such as $\mu\to 3 e$ and $\mu N\to e N$. 
We use a model independent approach in this analysis, where these processes are considered to be loop induced by exchanging spin-1/2 and spin-0 particles.
We explore two complementary cases, which has no or has an internal (built-in) cancellation mechanism in amplitudes.
Our main results are as following.
(a) Bounds from rates are used to constrain parameters, such as coupling constants and masses. These constraints can be easily updated by simple scalings, if the experimental situations change.
(b) The muon $g-2$ data favors non-chiral interactions.
(c) In $\mu\to 3 e$ and $\mu N\to e N$ processes, $Z$-penguin diagrams may play some role, while box diagrams contributions to $\mu\to3e$ are usually highly constrained.
(d) In the first case (without any built-in cancellation mechanism), using the recent $\mu\to e\gamma$ bound, we find that $\mu \to 3e$ and $\mu N\to e N$ rates are usually bounded below the present experimental limits by two to three orders of magnitudes in general.  Furthermore, by comparing $\Delta a_\mu$ and ${\cal B}(\mu\to e\gamma)$ data, the couplings of $\mu$ and $e$ are found to be highly hierarchical. Additional suppression mechanism should be called for.
(e) In the second case (with a built-in cancellation mechanism), 
mixing angles can provide additional suppression factors to satisfy the $\Delta a_\mu$ and ${\cal B}(\mu\to e\gamma)$ bounds.
While the $\mu\to 3 e$ rate remains suppressed, 
the bounds on $\mu N\to e N$ rates, implied from the latest $\mu\to e\gamma$ bound, can be relaxed significantly and can be just below the present experimental limits. 
\bigskip
\small

\pacs{Valid PACS appear here}

\maketitle

%
%

\section{Introduction}

Charge lepton flavor violating (LFV) processes are prohibited in the Standard Model (SM) and, hence, are excellent probes of New Physics (NP).
Recently the search of $\mu\to e\gamma$ decay was reported by MEG collaboration giving~\cite{MEG}
\be
{\cal B}(\mu^+\to e^+\gamma)\leq 2.4\times 10^{-12}.
\en
The bound is several times lower than the previous one~\cite{PDG}. 
This result received a lot attentions (see, for example, \cite{after MEG1, after MEG2, after MEG3}).
In many New Physics models this decay mode is closely related to other lepton flavor violating processes, such as $\mu^+\to e^+e^+e^-$ decays and $\mu^- N\to e^- N$ conversions~\cite{review}.
The present limits and future experimental sensitivities~\cite{MEG,PDG,future} of these LFV processes are summarized in Table~\ref{tab:expt bounds}.
Note that present bounds on $\mu$ LFV rates are roughly of similar orders.
It will be interesting to see what are the implications of the new ${\cal B}(\mu\to e\gamma)$ bound on these LFV processes and the interplay between them.

Since 2001, the muon anomalous magnetic moment remains as a
hint of a NP contribution (see, for a review, \cite{g-2 report}).
Experimental data deviates from the Standard Model (SM) expectation by more than 3$\sigma$~\cite{PDG}:
\be
\Delta a_{\mu}=a^{\rm exp}_\mu-a^{\rm SM}_\mu=(287\pm63\pm49)\pm 10^{-11}.
\en
Since NP contributes to $\Delta a_\mu$ and ${\cal B}(\mu^+\to e^+\gamma)$ through very similar loop diagrams [see Fig.~1(a) and (b)], it is useful to compare them at the same time.

\begin{table}[b]
\caption{Current experimental upper limits and future sensitivities on various muonic LFV processes~\cite{MEG,PDG,future}.}
 \label{tab:expt bounds}
\begin{ruledtabular}
\begin{tabular}{ l c  c }
~~~~~~
    & current limit 
    & future sensitivity
    \\
    \hline
${\cal B}(\mu^+\to e^+\gamma)$
    & $<2.4\times 10^{-12}$
    &  $10^{-13}$
    \\    
${\cal B}(\mu^+\to e^+e^+e^-)$
    & $<1.0\times 10^{-12}$
    &  $10^{-14}-10^{-16}$
    \\ 
${\cal B}(\mu^- {\rm Ti}\to e^-{\rm Ti})$
    & $<4.3\times 10^{-12}$
    & $10^{-18}$ 
    \\   
${\cal B}(\mu^- {\rm Au}\to e^-{\rm Au})$
    & $<7\times 10^{-13}$
    & $10^{-14}-10^{-16}$
    \\ 
${\cal B}(\mu^- {\rm Al}\to e^-{\rm Al})$
    & $\cdots$
    & $10^{-16}$
    \\                 
\end{tabular}
\end{ruledtabular}
\end{table}

The Large Hadron Collider (LHC) is working well. So far no NP signal is found (see, for example~\cite{LHC}). 
Plenty of well studied NP models or scenarios are ruled out or cornered.  
Therefore,  it will be useful to study low energy effect, when the NP scale is still beyond our reach. 
Given the present status on NP models, we believe that it is worthy to use a model independent approach.

In this work we consider a class of models that muon $g-2$ and various muon lepton flavor violating processes, such as $\mu\to e \gamma$, $\mu\to3e$ and $\mu\to e$ conversions, are loop-induced by exchanging spin-1/2 and spin-0 particles.
We try to see where the present $g-2$ and $\mu\to e \gamma$ experimental results lead us to on estimating rates or bounds on various LFV muonic decay modes and the interplay between them. 
Two cases, which are complementary to each other, are considered. In the first case, there is no any built-in cancellation mechanism among amplitudes. 
The second case is with some built-in mechanism, such as Glashow-Iliopoulos-Maiani (GIM) or super GIM mechanism. 
These two cases will be compared.

The lay out of this work is as following. In the next section, the framework is given. Numerical results are presented in Sec. III, where bounds from rates are used to constrain parameters, such as coupling constants and masses. Correlations between different processes are investigated.
Discussion and conclusion are given in Sec. IV and V, respectively. 
Some formulas and additional informations are collected in Appendices.

\section{Framework}

\begin{figure}[t]
\centering
\subfigure[]{
  \includegraphics[width=6cm]{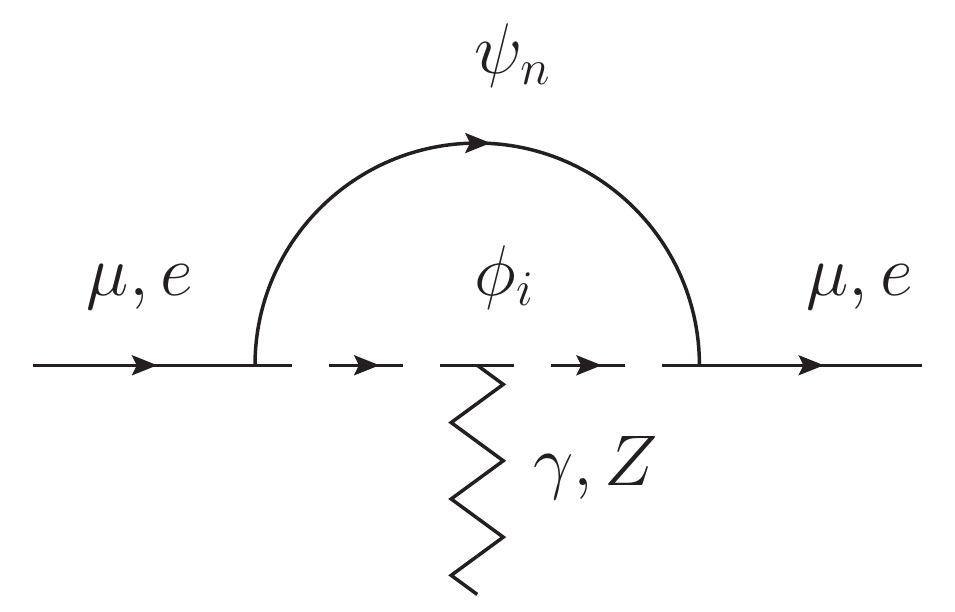}
}
\subfigure[]{
  \includegraphics[width=6cm]{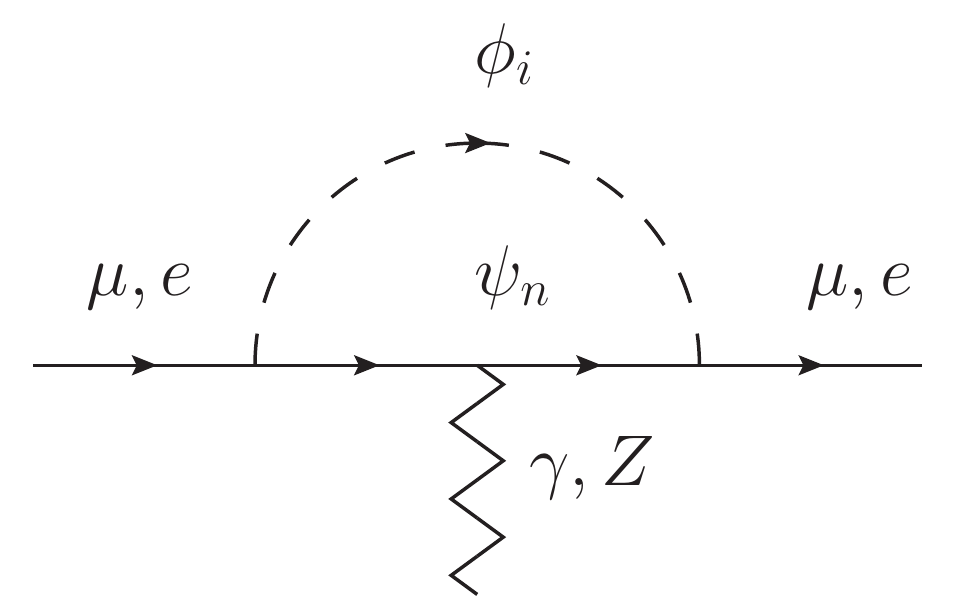}
}
\subfigure[]{
  \includegraphics[width=6cm]{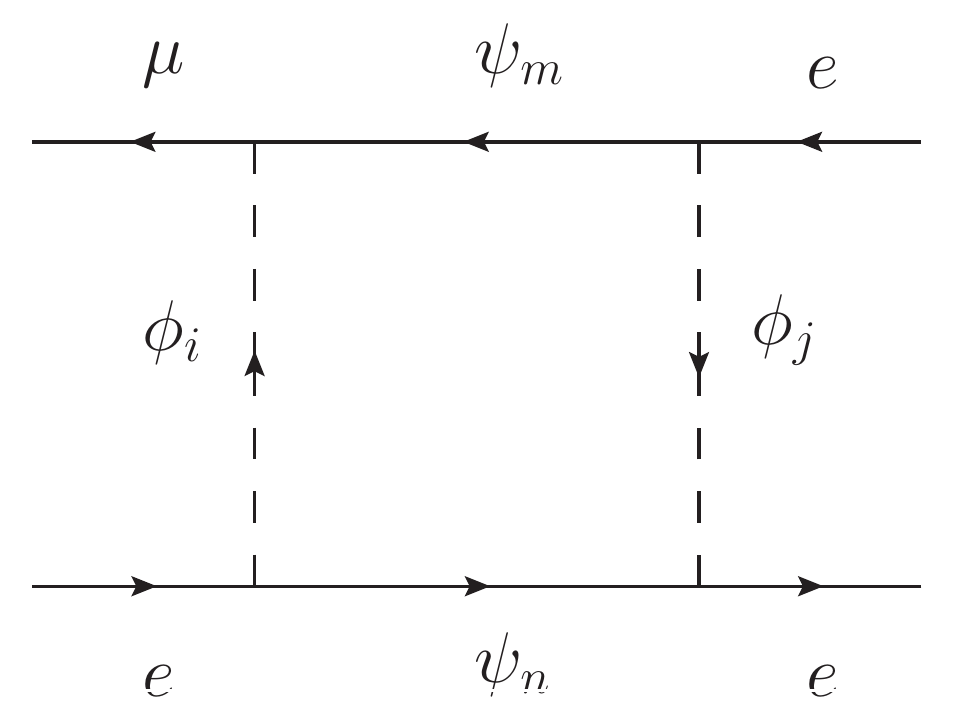}
}
\subfigure[]{
  \includegraphics[width=6cm]{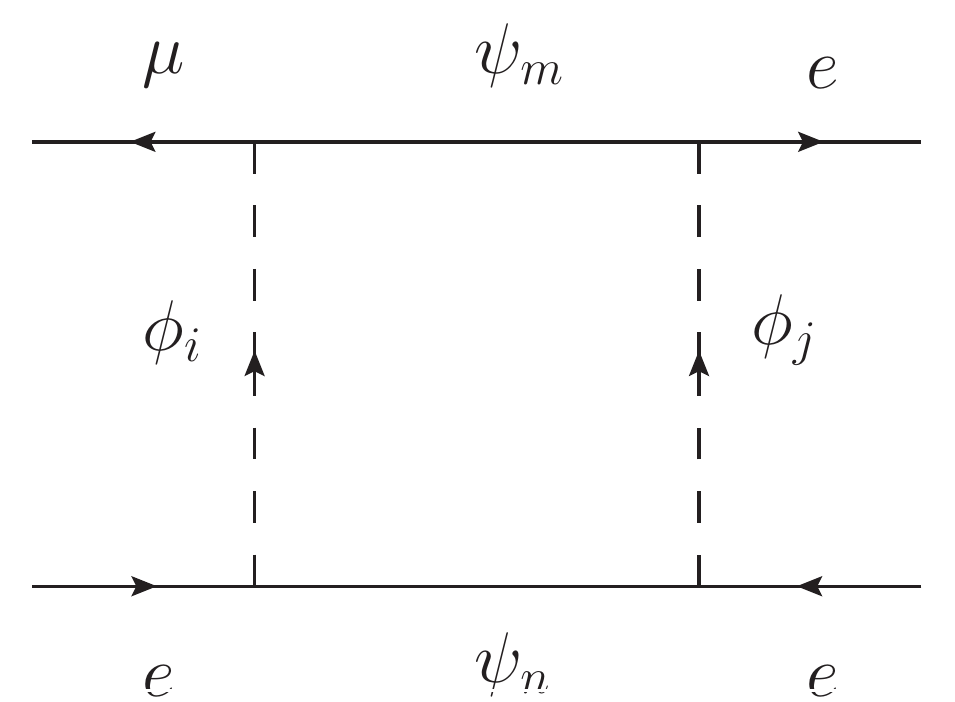}
}
\caption{(a) and (b): Penguin diagrams that contribute to muon $g-2$, $\mu^+\to e^+\gamma$, $\mu^+\to e^+ e^+ e^-$ and $\mu^- N\to e^- N$ processes. Note that diagrams involving self energy parts are not shown.
(c) and (d): Box diagrams contributing to the $\mu^+\to 3 e$ process. Figure (d) takes place only when $\psi_{m,n}$ are Majorana fermions.}
\label{fig:penguin&box}
\end{figure}

In this section, we begin with introducing the lagrangian of
a generic interaction involving leptons, exotic spin-1/2 fermions and spin-0 bosons.
Formulas of processes of interested, discussions on subtleties on the calculation of the $Z$-penguin amplitude and explicit expressions of Wilson coefficients will be given subsequently.  
This section end after the formulation of the two complementary cases as briefly mentioned in Sec. I.

\subsection{The interacting lagrangian and diagrams}\label{sec. Lint}

The lagrangian of a generic interaction involving leptons ($l$), exotic spin-1/2 fermions ($\psi_n$) and spin-0 bosons ($\phi_i$) is given by
\be
{\cal L}_{\rm int}=\bar\psi_n(g^{ni}_{lL} P_L+g^{ni}_{lR} P_R) l\phi_i^*
+\bar l(g_{lL}^{ni*}P_R+g_{lR}^{ni*}P_L)\psi_n\phi_i,
\label{eq:Lint}
\en
where summation over indices are understood unless specified.
The lagrangian is given in the mass bases and are ready to be used in calculations.
However, it is important to make sure that it transforms as a singlet under the SM gauge transformation.
 
In the weak bases of $\psi_{L p}$, $\psi_{R p}$, $\phi_{L a}$ and $\phi_{R a}$, the interacting lagrangian is
\be
{\cal L}_{\rm int}= 
(g'{}^{pa}_{lL} \bar\psi_{R p} l_L\phi_{L a}^*+g'{}^{pa}_{lR} \bar \psi_{L p} l_R\phi_{R a}^*)
+h. c.,
\label{eq:Lint weak}
\en
where we denote $\phi_{L(R)}$ for the scalar fields that couple to $l_{L(R)}$ and subscripts $p$ and $a$ are the labels of different weak fields, 
which may have different SU(2)$\times$U(1) quantum numbers.   
It is important to require that ${\cal L}_{\rm int}$ transforms as a singlet under the SM gauge groups and quantum numbers of $\psi$ and $\phi$ are related (see Appendix~\ref{app: QN}).

The mass bases are related to the weak bases through the following transformations
\be
\phi_i=U^L_{ia} \phi_{L a}+U^R_{i a}\phi_{R a},
\quad
\psi_{nL(R)}=V^{L(R)}_{np} \psi_{L(R) p},
\en
where $i$ and $n$ are labels of mass eigenstates and $U$ and $V$ are the mixing matrices relating weak and mass eigenstates.
With
\be
g^{ni}_{lL(R)}=g'{}^{pa}_{lL(R)} V^{R(L)}_{np} U^{L(R)}_{ia}
\en
the interacting lagrangian is now brought into the form shown in Eq. (\ref{eq:Lint}), which is more convenient and will be used in later calculations.

These interactions will induce lepton flavor violating processes at one-loop level.
Penguin diagrams contributing to the muon anomalous magnetic moment and $\mu^+\to e^+\gamma$ are shown in Fig.~\ref{fig:penguin&box}(a) and (b),
while box diagrams contributing to the $\mu^+\to 3 e$ process are shown in Fig.~\ref{fig:penguin&box}(c) and (d).
Note that (i) penguin diagrams shown in Fig.~\ref{fig:penguin&box}(a) and (b) also contribute to $\mu^+\to 3 e$ and $\mu N\to e N$ process by connecting the virtual photons or the $Z$ bosons to electron currents and quark currents, respectively,~\footnote{ 
It is possible to have box diagrams with the electron line in Fig.~\ref{fig:penguin&box}(c), (d) replaced by a quark one and contribute to $\mu N\to e N$ conversions in some cases. 
We will consider this contribution in the discussion section.}  (ii) Fig.~\ref{fig:penguin&box}(d) takes place only when $\psi_{m,n}$ are Majorana fermions.

\subsection{Formulas for various processes}\label{app:Formula}

To define our variables and to specify our convention, we collect formulas for various precesses here.
The relevant effective lagrangian in this study is
\be
{\cal L}_{\rm eff}={\cal L}_{l'l\gamma}+{\cal L}_{l'lll}+{\cal L}_{l'lqq}
\en
with $l^{(\prime)}=e,\mu,\tau$ and $q$ denoting quarks.
Each term will be specified in below.
For $l'\neq l$, we have
\be
\label{eq:dipole}
{\cal L}_{l'l\gamma}=\bar l'_L\sigma_{\mu\nu} l_R F^{\mu\nu} A_{L' R}
                              +\bar l'_R\sigma_{\mu\nu} l_L F^{\mu\nu} A_{R' L}
                              +h.c.,
\en
and
\be
A_{L R'}=A^*_{R' L},
\quad
A_{R L'}=A^*_{L' R}.
\en
Note that in the case of $l'=l$, we do not need the additional hermitian conjugated terms in Eq.~(\ref{eq:dipole}).
These $A$s are from the so-called $F_2$ photonic penguin and their explicit forms will be given later.

The effective lagrangians for $\bar l'\to 3 l$ decays and $l'\to l$ conversions are~\cite{review}
\be
{\cal L}_{l'lll}
&=&g_{R'LRL}(\bar {l'}_R l_L)(\bar l_R l_L)
        +g_{L'RLR}(\bar {l'}_L l_R)(\bar l_L l_R)
\non\\ 
&&        +g_{R'RRR}(\bar{l'}_R\gamma^{\mu} l_R)(\bar l_R\gamma_\mu l _R)
        +g_{L'LLL}(\bar{l'}_L\gamma^{\mu} l_L)(\bar l_L\gamma_{\mu} l_L)
\non\\        
&&  +g_{R'RLL}(\bar{l'}_R\gamma^{\mu} l_R)(\bar l_L\gamma_{\mu} l_L)      
        +g_{L'LRR}(\bar{l'}_L\gamma^{\mu} l_L)(\bar l_R\gamma_{\mu} l_R)
        +h.c.,
\\
{\cal L}_{l'lqq}
&=&\sum_{q=u,d}
 [g_{LV}(q)\bar{l'}_L\gamma^{\mu} l_L+g_{RV}(q)\bar{l'}_R\gamma^{\mu} l_R]\bar q\gamma_{\mu} q
   +h.c., 
\label{eq: Leff}   
\en        
where 
\be
g_{M'MNO}&\equiv&e^2 Q_l g^P_{M'M}\delta_{NO}+g^Z_{M'M} g^Z_{l_N}\delta_{NO}+g^B_{M'MNO},
\non\\
g_{MV}(q)&=&e Q_q^2 g^P_{M'M}+\frac{1}{2}g^Z_{M'M}(g^Z_{q_L}+g^Z_{q_R}),
\label{eq:effective}  
\en
for $M$, $N$, $O$=$L$, $R$ with $g^P_{M'M}$ from the so-called $F_1$ photonic penguin, $g^Z_{M'M}$ from the $Z$-penguin and $g^B_{M'MNN}$ from box diagrams.
More details and the explicit forms of these Wilson coefficients 
will be given later.
Note that although the above ${\cal L}_{l'lqq}$ is not the most generic one, it contains all the relevant parts that 
are closely related to ${\cal L}_{l'lll}$ and ${\cal L}_{l'l\gamma}$.

We now collect the formulas for various processes originated from the above Lagrangians. Comparing the effective lagrangians of the lepton $g-2$ and the electric dipole moment (EDM),~\footnote{We use the convention where $D_\mu=\partial_\mu+i eQA_\mu$ with $e=+|e|$.}
\be
{\cal L}_{g-2}=-\frac{e Q}{4 m_l} \Delta a_l\,\bar l\sigma_{\mu\nu}l F^{\mu\nu},
\quad
{\cal L}_{EDM}=-\frac{i}{2} d_l\,\bar l\sigma_{\mu\nu}\gamma_5 l F^{\mu\nu},
\en
to the generic expressions in Eq.~(\ref{eq:dipole}), the anomalous magnetic moment and
EDM of lepton $l$ can be readily read off as
\be
\Delta a_l=-\frac{4 m_l}{e Q_l} {\rm Re}(A_{R L}),
\quad
d_l=2{\rm Im}( A_{RL}),
\en
respectively.
The $\bar l'\to \bar l\gamma$ and $\bar l'\to \bar l\bar l l$ decay rates are given by
\be
\Gamma(\bar l'\to \bar l\gamma)=\frac{(m_{l'}^2-m_l^2)^3}{4\pi m_{l'}^3}
\left(|A_{L'R}|^2+|A_{R'L}|^2\right)
\en
and~\cite{review}
\be
\Gamma(\bar l'\to \bar l\,\bar l\, l)
&=&\frac{m_{l'}^5}{3(8\pi)^3}\Bigg[\frac{|g_{R'LRL}|^2}{8}+|g_{R'RLL}|^2+
32\left|\frac{e A_{R'L}}{m_{l'}}\right|^2\log (\frac{m^2_{l'}}{m^2_l}-\frac{11}{4})
\non\\
&&
+16 {\rm Re}\left(\frac{e A_{R'L} g^*_{L'LLL}}{m_{l'}}\right)
+8{\rm Re}\left(\frac{e A_{R'L} g^*_{L'LRR}}{m_{l'}}\right)\Bigg]
\non\\
&&
+L\leftrightarrow R,
\en
respectively. While the $l' N \to l N$ conversion rate ratio is governed by 
\be
{\cal B}_{l'N\to eN}=\frac{\omega_{\rm conv}}{\omega_{\rm capt}},
\en
with
\be
\omega_{\rm conv}&=&
\left|\frac{A^*_{R'L} D}{2 m_{l'}}+2 [2 g^*_{LV}(u)+g^*_{LV}(d)] V^{(p)}
+2 [g^*_{LV}(u)+2g^*_{LV}(d)] V^{(n)}\right|^2
+L\leftrightarrow R.
\label{eq: conv}
\en
The numerical values of $D$, $V$ and $\omega_{\rm capt}$ are taken from \cite{KKO,capt} and are collected in Appendix~\ref{app:FG}.

\subsection{$Z$-penguin amplitudes}\label{subsec:Z}

The calculation of the $Z$-penguin amplitude is quite complicate and subtle. Some explanations are needed.

The interaction involving a $Z$ boson is given by
\be
{\cal L}^Z_{\rm int}&=&-\bar l\Zslash(g_{l_L}^Z P_L+g^Z_{l_R} P_R)l
                                -\bar\psi_{L p}\Zslash g_{\psi_{L p}}^Z\psi_{L p}
                                -\bar\psi_{R q}\Zslash g^Z_{\psi_{R q}} \psi_{R q} 
\non\\
         &&                       -ig^Z_{\phi_{L a}}(\phi_{L a}^*\partial^\mu\phi_{L a}-\partial^\mu\phi_{L a}^*\phi_{L a})
                                -ig^Z_{\phi_{R a}}(\phi_{R a}^*\partial^\mu\phi_{R a}-\partial^\mu\phi_{R a}^*\phi_{R a})+\dots
\en
with
\be
g^Z_X=\frac{e}{\sin\theta_W\cos\theta_W}(T_3-\sin^2\theta_W Q)_X,
\label{eq:gZ}
\en
for $X=l_{L(R)}$, $\psi_{L(R)p}$ and $\phi_{L(R)a}$ in the weak eigenstates.
Since ${\cal L}_{\rm int}$ transforms as a singlet under the SM gauge group, the $g^Z_X$ of various fields are related through \be
g_{l_{L(R)}}^Z-g^Z_{\psi_{R(L)}\, p}-g^Z_{\phi\,a}=0,
\label{eq: gZ relation}
\en
if the corresponding coupling $g'{}^{pa}_{lL(R)}$ in Eq.~(\ref{eq:Lint weak}) is  non-vanishing.

Although the couplings $g^Z_{\psi_{L(R)},\phi_{L(R)}}$ are diagonal in the weak bases, it may have off-diagonal terms in the mass bases. 
In the mass bases, the interacting lagrangian involving a $Z$ boson is given by
\be
{\cal L}^Z_{\rm int}&=&-\bar l\Zslash(g_{l_L}^Z P_L+g^Z_{l_R} P_R)l
                                -\bar\psi_m\Zslash (g_{\psi_L\, mn}^Z P_L+g_{\psi_R\, mn}^Z P_R)\psi_{n}
                               \non\\
         &&                       -ig^Z_{\phi\,ij}(\phi_i^*\partial^\mu\phi_j-\partial^\mu\phi_i^*\phi_j)
                              +\dots
\en
with
\be
g^Z_{\phi\, ij}=U^L_{i a} g^Z_{\phi_{L a}} U^{\dagger L}_{a j}+U^R_{i a} g^Z_{\phi_{R a}} U^{\dagger R}_{a j},
\quad
g^Z_{\psi_{L(R)}\, mn}=V^{L(R)}_{mp} g^Z_{\psi_{L(R) p}} V^{\dagger L(R)}_{pn}.
\en

The one-loop amplitude for $l\to l'Z$ consists of two diagrams shown in Fig.~1 
and two additional diagrams involving self-energy diagrams with $Z$ attached to external lines. 
The resulting amplitude is given by (neglecting $m_l$, $m_{l'}$ and $q^2$)
\be
iM&=&\frac{i}{16\pi^2}\bar u'(g^{mi*}_{l' L} P_R+g^{mi*}_{l' R} P_L)
\eslash^*
[(g_{l_L}^Z\delta_{ij}\delta_{mn}-g^Z_{\psi_R\, mn}\delta_{ij}-g^Z_{\phi\,ij}\delta_{mn})P_L
\non\\
&&
+(g_{l_R}^Z\delta_{ij}\delta_{mn}-g^Z_{\psi_L\,mn}\delta_{ij}-g^Z_{\phi\,ij}\delta_{mn})P_R]
(g^{nj}_{l L} P_L+g^{nj}_{l R} P_R)u
\non\\
&&\times\left[\frac{1}{2}\left(\frac{2}{4-d}-\gamma_E+\ln\frac{4\pi}{M^2}\right)-
F_Z(m_{\psi_m}^2, m_{\psi_n}^2, m_{\phi_i}^2, m_{\phi_j}^2, M^2)\right]
\non\\
&&
+\frac{i}{16\pi^2}(g_{\psi_R\,mn}^Z-g^Z_{\psi_L\, mn})\bar u'(g^{mi*}_{l' L} P_R+g^{mi*}_{l' R} P_L)
\eslash^*
(g^{ni}_{l L} P_L-g^{ni}_{l R} P_R)u\,G_Z(m_{\psi_m}^2, m_{\psi_n}^2, m_{\phi_i}^2),
\non\\
\en
where $d$ is the number of the space-time dimension, $\gamma_E$ is the Eular number and $F_Z$ and  $G_Z$ are the loop functions whose explicit forms shown in Appendix~\ref{app:FG}. Note that $M$ is an arbitrary mass parameter introduced to balance dimension. We will return to it later.

Note that the divergent part contained in the first term is indeed vanishing by requiring the $l$--$\psi$--$\phi$ interaction
${\cal L}_{\rm int}$ in Eq.~(\ref{eq:Lint weak}) be invariant under the SM gauge group, i.e. we have
\be
g^{mi*}_{l' L(R)}(g_{l_{L(R)}}^Z\delta_{ij}\delta_{mn}-g^Z_{\psi_{R(L)}\, mn}\delta_{ij}-g^Z_{\phi\,ij}\delta_{mn})g^{nj}_{l L(R)}
=
g'{}^{pa*}_{l' L(R)}(g_{l_{L(R)}}^Z-g^Z_{\psi_{R(L)}\, p}-g^Z_{\phi\,a})g'{}^{pa}_{l L(R)}
=0,
\non\\
\en
where sum over indices are understood and Eq.~(\ref{eq: gZ relation}) has been used in the last step. 
The non-divergent part, namely the one with $F_Z$, survives. 
Note that by the same token the dependence on the arbitrary mass parameter $M$ cancels. 
The resulting Wilson coefficients will be given later. 
It is easy to see that in the non-mixing case ($U=V={\bf 1}$), the whole first term (including the $F_Z$ term) is vanishing.

As a cross check, we note that the same expression of $iM$ can be use to obtain the lowest order $\gamma$-penguin amplitude by replacing each $g^Z$ by the corresponding $e Q$. Since under these replacement 
$(g_{l_{L(R)}}^Z\delta_{ij}\delta_{mn}-g^Z_{\psi_{R(L)}\, mn}\delta_{ij}-g^Z_{\phi\,ij}\delta_{mn})\to
e (Q_l-Q_\psi-Q_\phi)\delta_{ij}\delta_{mn}=0$ and
$(g^Z_{\psi_L mn}-g^Z_{\psi_R mn})\to e (Q_\psi-Q_\psi)\delta_{mn}=0$, the corresponding $\bar u'\eslash^* u$ term is vanishing as expected.

\subsection{Wilson coefficients}\label{subsec:Wilson}

Induced by the interaction given in Eq.~(\ref{eq:Lint}) the Wilson coefficients for the effective lagrangian in the Sec.~II~A are calculated to be
\be
A_{M' N}&=&\frac{e}{32\pi^2}
[(m_l g_{l'M}^{ni*} g^{ni}_{lM}+m_{l'} g^{ni*}_{l'N} g^{ni}_{lN})(Q_{\phi_i} F_1(m_{\psi_n}^2,m_{\phi_i}^2)-Q_{\psi_n} F_1(m_{\phi_i}^2,m_{\psi_n}^2))
\non\\
&&+m_{\psi_n} g_{l'M}^{ni*} g^{ni}_{lN} (Q_{\phi_i} F_3(m_{\psi_n}^2,m_{\phi_i}^2)-Q_{\psi_n} F_2(m_{\phi_i}^2,m_{\psi_n}^2))],
\en
for $M\neq N$, but with $M,N=L,R$, and $F_i$ are loop functions collected in Appendix~\ref{app:FG}.
The Wilson coefficients in Eq.~(\ref{eq:effective}) are
\be
g^P_{R'R}&=&\frac{1}{16\pi^2}\{g^{ni*}_{l'R} g^{ni}_{lR} 
                                             [Q_{\psi_n} G_2(m^2_{\phi_i},m^2_{\psi_n})+Q_{\phi_i} 
                                             G_1(m^2_{\psi_n},m^2_{\phi_i})]
\non\\
           &&  + m_{\psi_n}(m_l g^{ni*}_{l'R} g^{ni}_{lL}+m_{l'} g^{ni*}_{l'L} g^{ni}_{lR})
                                             [Q_{\psi_n} G_3(m^2_{\phi_i},m^2_{\psi_n})
                                             +Q_{\phi_i} G_3(m^2_{\psi_n},m^2_{\phi_i})]\}                                          
\non\\
g^Z_{R'R}&=&-\frac{1}{16\pi^2 m_Z^2\sin2\theta_W}2\kappa_{R\,ijmn}
                                             g^{mi*}_{l'R} g^{nj}_{lR}
                                             F_Z(m^2_{\psi_m},m^2_{\psi_n},m^2_{\phi_i},m^2_{\phi_j},m^2_Z)
                                             \non\\
                                             &&
-\frac{e}{16\pi^2 m_Z^2\sin2\theta_W}2 \Delta T_{3\psi mn}
                                             g^{mi*}_{l'R} g^{ni}_{lR}
                                             G_Z(m^2_{\psi_m},m^2_{\psi_n},m^2_{\phi_i}),
\non\\
g^Z_{L'L}&=&-\frac{1}{16\pi^2 m_Z^2\sin2\theta_W}2\kappa_{L\,ijmn}
                                             g^{mi*}_{l'L} g^{nj}_{lL}
                                             F_Z(m^2_{\psi_m},m^2_{\psi_n},m^2_{\phi_i},m^2_{\phi_j},m^2_Z)
                                             \non\\
                                             &&
                                             +\frac{e}{16\pi^2 m_Z^2\sin2\theta_W}2 \Delta T_{3\psi mn}
                                             g^{mi*}_{l'L} g^{ni}_{lL} 
                                             G_Z(m^2_{\psi_m},m^2_{\psi_n},m^2_{\phi_i}),
\non\\
g^B_{R'LRL}&=&\frac{1}{16\pi^2} F(m^2_{\psi_m},m^2_{\psi_n},m^2_{\phi_i},m^2_{\phi_j}) 
                           (g_{l' R}^{mi^*} g_{lL}^{mj} g_{lR}^{nj*} g_{l L}^{ni} 
                           -2\eta g_{l' R}^{mi^*} g_{lR}^{mj*} g_{l L}^{ni}  g_{lL}^{nj}), 
\non\\
g^B_{R'RRR}&=&\frac{1}{16\pi^2}
                                \bigg [\frac{\eta}{2}g_{l' R}^{mi*} g_{lR}^{mj*} g_{l R}^{ni} g_{lR}^{nj}
                                 F(m^2_{\psi_m},m^2_{\psi_n},m^2_{\phi_i},m^2_{\phi_j}) 
\non\\                                 
 &&                          -\frac{1}{4}g_{l' R}^{mi*} g_{lR}^{mj} g_{lR}^{nj*} g_{l R}^{ni}
                                 G(m^2_{\psi_m},m^2_{\psi_n},m^2_{\phi_i},m^2_{\phi_j})\bigg] , 
\non\\
g^B_{R'RLL}&=&\frac{1}{16\pi^2}\bigg\{ 
                            -\frac{1}{4}G(m^2_{\psi_m}, m^2_{\psi_n},m^2_{\phi_i},m^2_{\phi_j})
                            (g_{l' R}^{mi^*} g_{lR}^{mj} g_{lL}^{nj*} g_{l L}^{ni} 
                            +\eta g_{l' R}^{mi^*} g_{lL}^{mj*} g_{l L}^{ni} g_{lR}^{nj}) 
\non\\
 &&         -\frac{1}{2}g_{l' R}^{mi*} g_{lL}^{mj}  g_{lL}^{nj*} g_{l R}^{ni}
                                    F(m^2_{\psi_m}, m^2_{\psi_n},m^2_{\phi_i},m^2_{\phi_j}) 
\non\\          
 &&                +\frac{\eta}{4} g_{l' R}^{mi*} g_{lL}^{mj*} g_{l R}^{ni} g_{lL}^{nj}
                           G(m^2_{\psi_m}, m^2_{\psi_n},m^2_{\phi_i},m^2_{\phi_j}) \bigg\}, 
\label{eq:gPgB}
\en
with 
\be
\kappa_{L(R)ijmn}&\equiv& \sin2\theta_W (g_{l_{L(R)}}^Z\delta_{ij}\delta_{mn}-g^Z_{\psi_{R(L)}\,mn}\delta_{ij}-g^Z_{\phi\,ij}\delta_{mn})/e, 
\non\\
\Delta T_{3\psi mn}&\equiv& V^R_{mp}T_{3\psi_R p} V^{\dagger L}_{pn}-V^L_{mp}T_{3\psi_L p} V^{\dagger R}_{pn},
\en 
loop functions $F$ and $G_{(i)}$ shown in the Appendix~\ref{app:FG} and $\eta=1(0)$ for Majorana (Dirac) fermionic $\psi$. Other $g$ can be obtained from the above ones by exchanging $R$ and $L$. Note that for definiteness we take $M=m_Z$ in $F_Z$.
As before the summation on $m,n,i,j$ is understood.

\subsection{Two Cases}

We consider two complementary cases.

\subsubsection{Case I}
In the first case, namely case I, there is no built-in cancellation mechanism. 
The amplitudes may contain $N$ different sub-amplitudes, each comes from one of the loop diagrams as shown in Fig.~\ref{fig:penguin&box},
\be
A=\sum_{j=1}^N A_j.
\label{eq: N A}
\en
We will constrain parameters from data by switching various diagrams (sub-amplitudes) on one at a time. 
The corresponding Wilson coefficients of a typical sub-amplitude can be obtained by using formulas in Sec.~\ref{subsec:Wilson}, but with the replacement, 
\be
g^{ni}_{lM}\to g_{lM},
\en
with all summation on $n$ and $l$ suspended.
Since there is no built-in cancellation in this case, different sub-amplitudes are in principle independent from each other. 
Although it is likely to have various amplitudes to appear at the same time in a realistic model calculation and to interfere with each others, the interference effects only become important if the amplitudes are of similar size. Hence, our analysis not only is valid when the sizes are different (hence, constraining the most dominant amplitude), but can also provide information on regions where interference may be important.

\subsubsection{Case II}

In case II, there is a built-in cancellation such as a GIM or a super-GIM mechanism in the NP sector. 
This case is complementary to the previous one.
Some of the sub-amplitudes in Eq.~(\ref{eq: N A}) are intimately related. We have to group them to allow the cancellation mechanism to do its job first. The grouped amplitudes should be viewed as new sub-amplitudes and we will turn them on one at a time to constrain their sizes from data.
To be specify, we consider
\be
g^{ni}_{lM}\to g^{i}_{lM}=g_{lM}\Gamma^{il}_M,
\en
where we have $M=L,R$ and $g_{lM}$ is real as the phase is absorbed into $\Gamma$.
Note that the matrix $\Gamma$ is similar and related to the mixing matrix $U$, but is not identical to it. 
These $\Gamma$ satisfy the following relations:
\be
\Gamma^{\dagger li}_M m^2_i\Gamma^{il'}_N=(m^2_\phi) ^{ll'}_{MN},
\quad
\Gamma^{\dagger li}_M\Gamma^{il'}_N=\delta^{ll'}\delta_{MN}.
\en
A typical expression of Wilson coefficients given in Sec.~\ref{subsec:Wilson} is transformed in the following way:
\be
\sum_i g^{i*}_{\mu M} f(m^2_{\psi}, m^2_{\phi_i}) g^{i}_{e N}
\to m^2_\phi\frac{\partial }{\partial m^2_\phi}f(m^2_{\psi}, m^2_{\phi}) g_{\mu M} g_{e N} \delta^{MN}_{\mu e},
\en
where $m^2_\phi$ is the average mass squared of $\phi_i$ and the mixing angle $\delta^{MN}_{\mu e}$ is defined in the usual way to be~\cite{Gabbiani:1996hi}
\be
\delta^{MN}_{\mu e}
\equiv \frac{1}{m^2_\phi}
\Gamma^{M\dagger}_{\mu i} (m^2_{\phi_i}-m^2_{\phi}) \Gamma^N_{i e}
=\frac{(m^{2 MN}_\phi)_{\mu e}}{m^2_\phi}.
\en
The Wilson coefficients in this case can be obtained readily by applying the above replacements to the generic formulas collected in Sec.~\ref{subsec:Wilson}.

Note that in the $Z$ penguin amplitude the zeroth and first order 
terms in the $\kappa_{L(R)} F_Z$ part are vanishing.
The leading order contribution is at the level of $\delta_{LR}\delta_{RL}$, which is beyond the accuracy of the present analysis and are neglected.  

\section{Results}

Numerical results in cases I and II are given in this section. 
Unless specified explicitly, experimental inputs are taken from Table~1 and Ref.~\cite{PDG}.

\subsection{Case I}\label{sec: case I}

\begin{figure}[t]
\centering
\subfigure[]{
  \includegraphics[width=7.5cm]{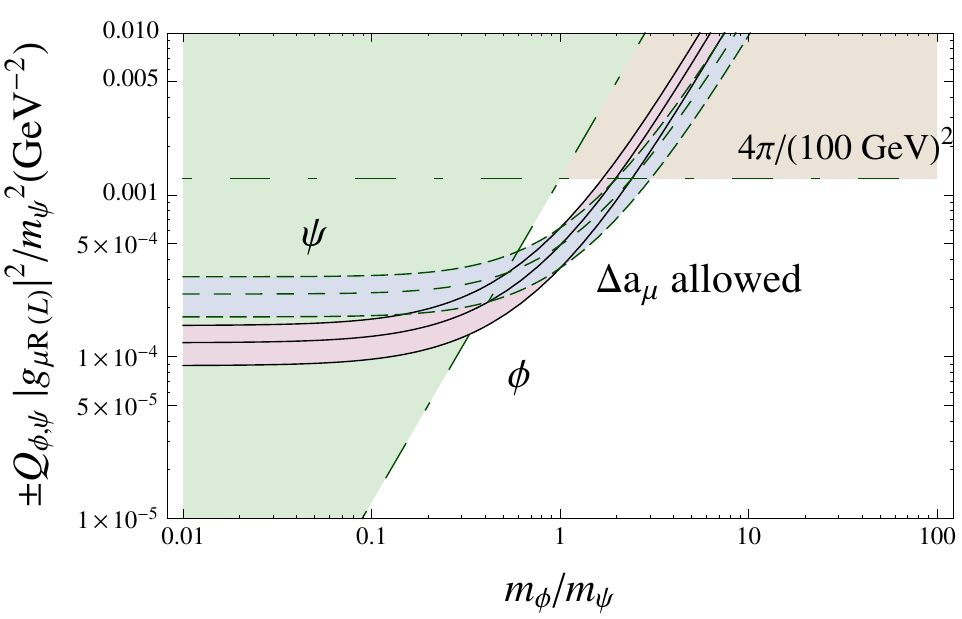}
}
\hspace{0.5cm}
\subfigure[]{
  \includegraphics[width=7.5cm]{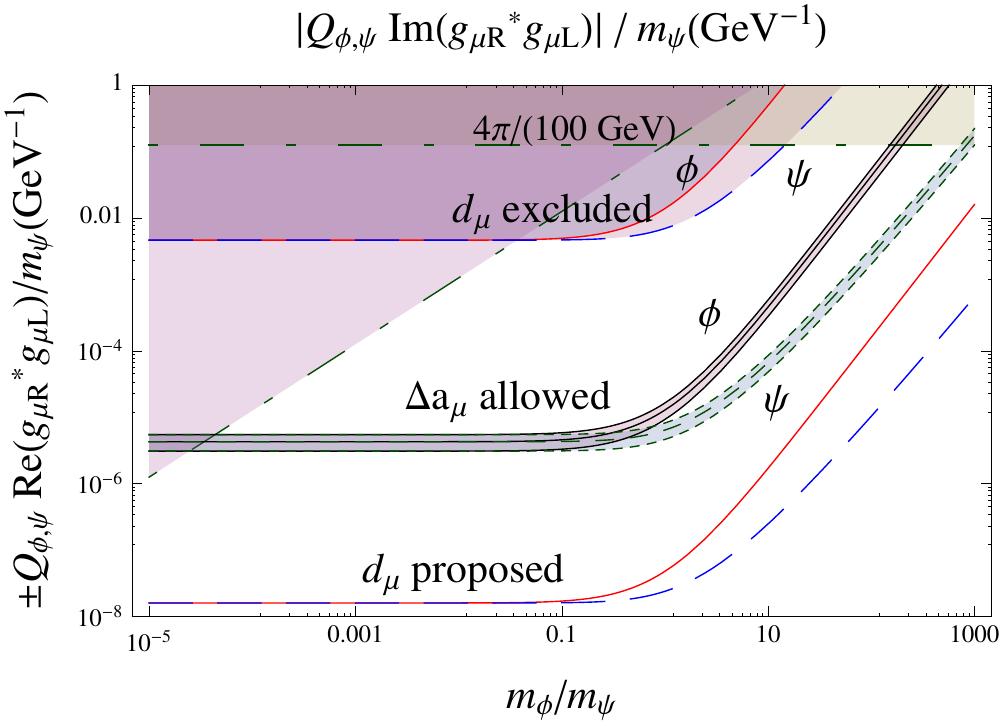}
}
\caption{(a) Allowed parameter space for $\pm Q_{\phi,\psi} |g_{\mu L(R)}|^2/m_\psi^2$ constrained by $\Delta a_\mu$ (bands with solid or dashed lines) with exclusion of $|g_{\mu L(R)}|^2>4\pi$ and $m_{\psi,\phi}<100$~GeV (shaded regions with dot-dashed lines).
(b) Allowed parameter space for $\pm Q_{\phi,\psi}{\rm  Re}(g^*_{\mu R} g_{\mu L})/m_\psi$ constrained by 
$\Delta a_\mu$ with exclusion of $|g_{\mu L} g_{\mu R}|>4\pi$ and $m_{\psi,\phi}>100$~GeV. Excluded parameter space (shaded regions with solid or dashed lines) of $|Q_{\phi,\psi} {\rm Im}(g^*_{\mu R} g_{\mu L})|/m_\psi$ from the muon EDM bound and the expected sensitivity are also shown.}
\label{fig:muon g-2 1}
\end{figure}

In Fig.~\ref{fig:muon g-2 1} we show the allowed parameter space for $Q_{\phi,\psi} |g_{\mu L(R)}|^2/m_\psi^2$ and $Q_{\phi,\psi}{\rm  Re}(g^*_{\mu R} g_{\mu L})/m_\psi$ constrained by the measured $\Delta a_\mu$ 
with exclusion of $|g_{\mu L(R)}|^2,|g_{\mu L} g_{\mu R}|>4\pi$ and $m_{\psi,\phi}<100$~GeV.~\footnote{
It is easy to see that $|g_{\mu L(R)}|^2>4\pi$ and $m_\psi<100$ GeV implies $|g_{\mu L(R)}|^2/m^{2}_\psi>4\pi/(100)^{2}$ GeV$^{-2}$ and $|g_{\mu L}g_{\mu R}|^2/m_\psi>4\pi/100$ GeV$^{-1}$, 
while $|g_{\mu L(R)}|^2>4\pi$ and $m_\phi<100$~GeV 
implies $|g_{\mu L(R)}|^2/m^{2}_\psi=(|g_{\mu L(R)}|^2/m^{2}_\phi)\times(m_\phi/m_\psi)^{2}
>4\pi/(100)^2\times(m_\phi/m_\psi)^{2}$~GeV$^{-2}$ and 
$|g_{\mu L} g_{\mu R}|^2/m_\psi
>(4\pi/100)\times(m_\phi/m_\psi)$~GeV$^{-1}$. 
These excluded regions are shown by shaded areas with horizontal or inclined boundaries.} 
The latter requirements are to ensure perturbativity and to satisfy the experimental bounds on the masses of exotic particles~\cite{PDG}. 
Bands denoted with $\phi$ or $\psi$ are allowed regions obtained through contributions from diagrams with $\phi$ or $\psi$ interacting with a photon [see Fig~\ref{fig:penguin&box}(a) and (b)].
Excluded parameter space of $|Q_{\phi,\psi} {\rm Im}(g^*_{\mu R} g_{\mu L})|/m_\psi$ confronting the muon electric dipole moment (EDM) bound~\cite{PDG} is also shown in Fig.~\ref{fig:muon g-2 1}(b).

From Fig.~\ref{fig:muon g-2 1}(a), we see that the allowed regions on $Q_{\phi} |g_{\mu L(R)}|^2/m_\psi^2$ and $-Q_{\psi} |g_{\mu L(R)}|^2/m_\psi^2$ are similar and the signs of $Q_{\phi,\psi}$ are constrained by data. 
Note that the allowed parameter space is quite limited. 
Indeed, it is almost closed by the bounds from $|g_{\mu L(R)}|^2<4\pi$ and $m_{\psi}>100$~GeV and $m_\phi>100$~GeV.
The allowed region is around $\pm Q_{\phi,\psi} |g_{\mu L(R)}|^2/m_\psi^2\simeq 10^{-4}\sim10^{-3}$~GeV$^{-2}$ and $m_\phi/m_{\psi}\simeq 0.3\sim 3$, 
which implies that for $m_{\phi,\psi}$ of a few hundred GeV, the couplings $Q_{\phi,\psi} |g_{\mu L(R)}|^2$ are required to be of order ${\cal O}(1)\sim{\cal O}(10)$, which are rather large, and are even larger for heavier $m_{\phi,\psi}$. 
To see it in another way, if we take the size of $g_{\mu L(R)}$ to be similar to that of the electric coupling $e$, 
we need $m_{\phi,\psi}$ to be as light as $10$ to 30 GeV to reproduce the experimental result on $\Delta a_\mu$. 
Thus, it is unlikely to use a chiral type interaction ($ g_{\mu L}\times g_{\mu R}=0$) to generate the measured $\Delta a_\mu$.

\begin{figure}[tb]
\centering
\subfigure[]{
  \includegraphics[width=7cm]{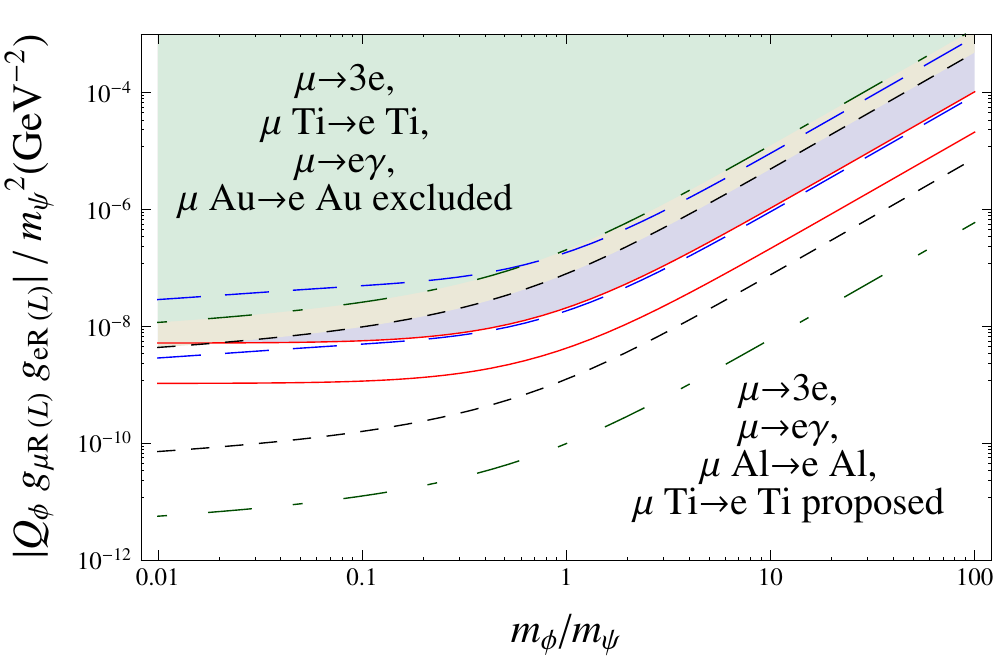}
}
\hspace{0.5cm}
\subfigure[]{
  \includegraphics[width=7cm]{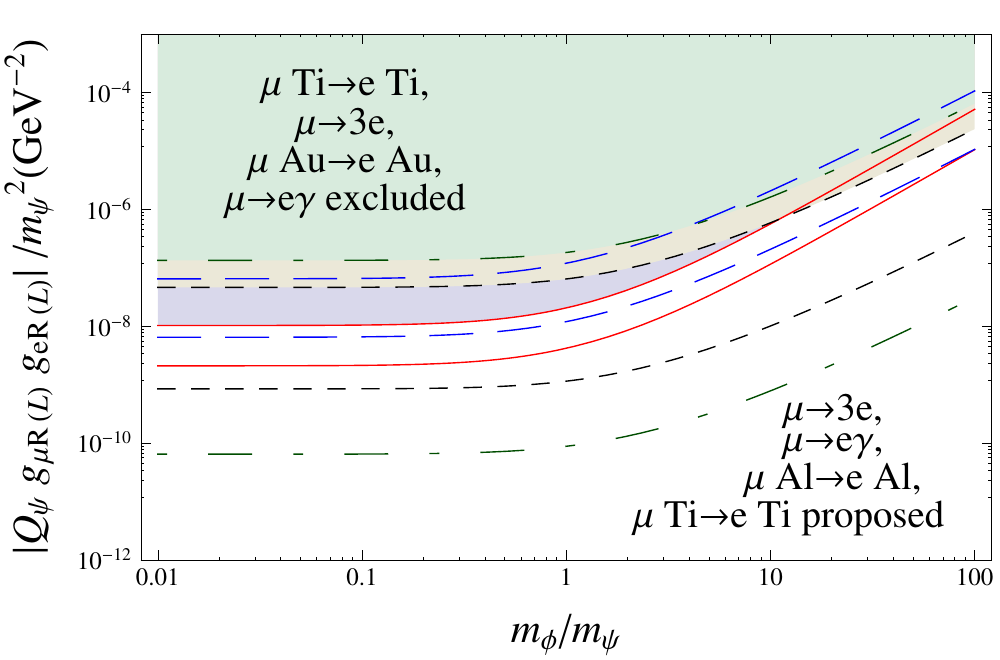}
}
\subfigure[]{
  \includegraphics[width=7cm]{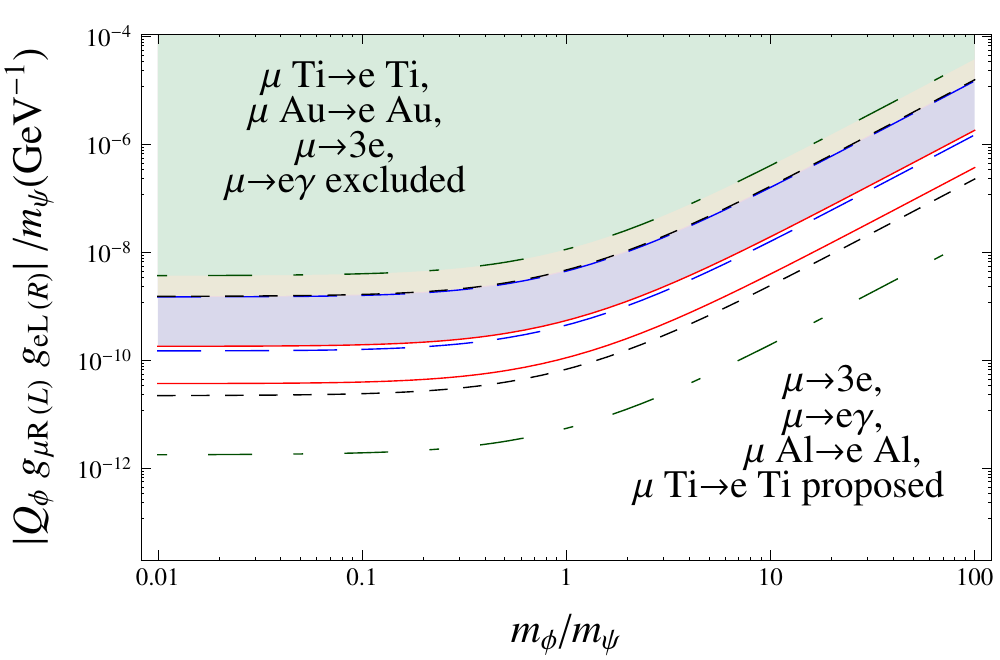}
}
\hspace{0.5cm}
\subfigure[]{
  \includegraphics[width=7cm]{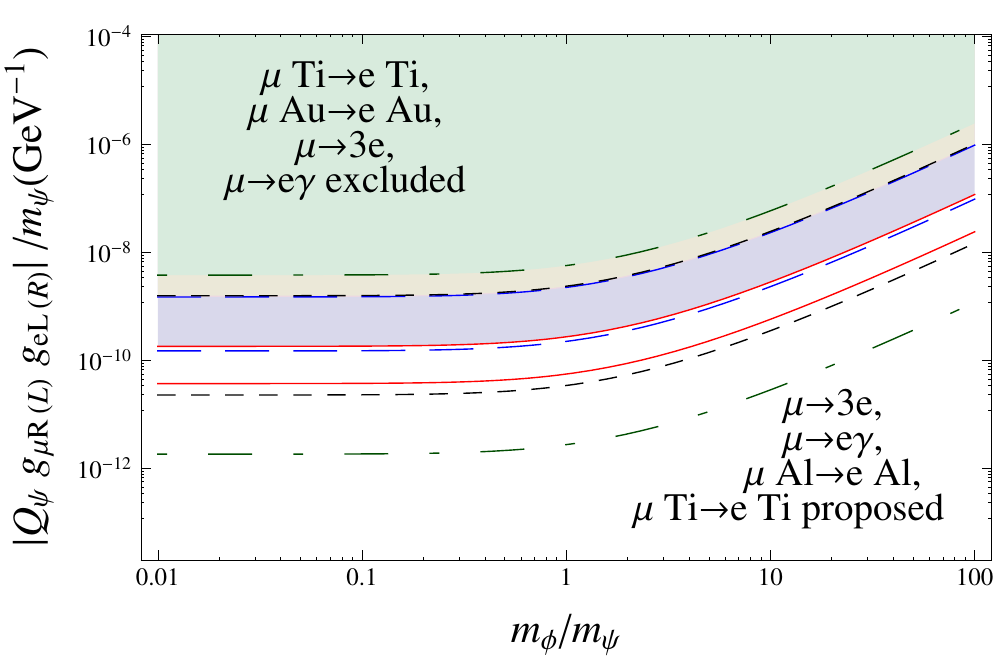}
}
\subfigure[]{
  \includegraphics[width=7cm]{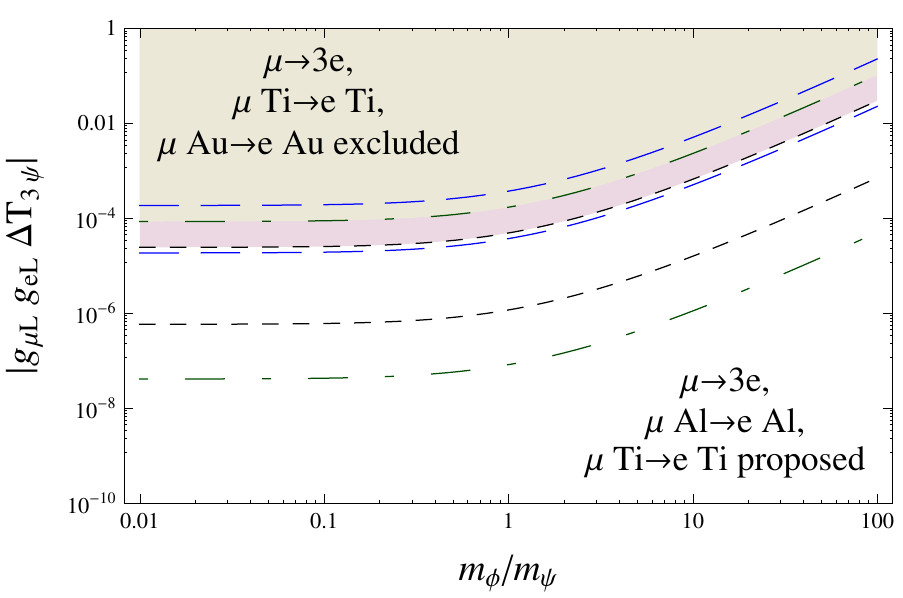}
}
\hspace{0.5cm}
\subfigure[]{
  \includegraphics[width=7cm]{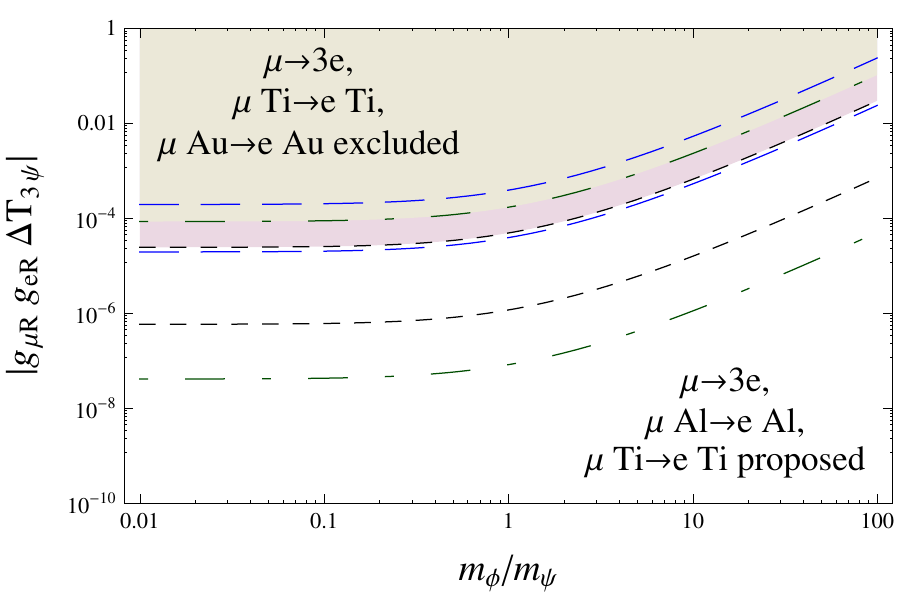}
}
\caption{(a)-(d): Parameter space excluded (projected) by various bounds (expecting sensitivities) on $\mu$ LFV processes through contributions from photonic penguins. Note that
solid, dashed, dot-dashed and short-dashed lines denote results from $\mu\to e \gamma$, $\mu\to 3 e$, $\mu {\rm Ti}\to e{\rm Ti}$ and $\mu {\rm Au(Al)}\to e {\rm Au(Al)}$ processes, respectively. (e) and (f): Same as (a)-(d), but through contributions from $Z$-penguins.}
\label{fig:LFVpenguin}
\end{figure}

From Fig.~\ref{fig:muon g-2 1}(b), we see that the allowed parameter space is substantially larger.
To reproduce the measured $\Delta a_\mu$, the mass ratio has to be in the range of ${\cal O}(10^{-5})\lesssim m_\phi/m_\psi\lesssim {\cal O}(10^{2,3})$, which is much wider than the one in Fig.~\ref{fig:muon g-2 1}(a). 
We note that the bands of the allowed parameter space behave rather differently in two regions roughly separated by $m_\phi/m_\psi= 0.1$.
(i)~For $m_\phi/m_\psi\lesssim0.1$, the horizontal bands denoting the allowed parameter region for $\pm Q_{\phi,\psi}{\rm  Re}(g^*_{\mu R} g_{\mu L})/m_\psi$ are around $4\times 10^{-6}$ GeV$^{-1}$. They are insensitive to $m_\phi/m_\psi$,
since the chiral enhancement factor $m_\psi/m_\mu$ compensates the suppression from the heavy $\psi$ mass. 
Note that $m_\phi/m_\psi$ can be as low as $3\times 10^{-5}$, which implies that $m_\psi$ up to 
$3\times10^3$~TeV 
is still capable to reproduce the measured $\Delta a_\mu$ in the extreme case, 
where we have a light $\phi$ [$m_\phi={\cal O}(100)$ GeV] and large couplings [$|g_{\mu L} g_{\mu R}|={\cal O}(4\pi) $].
(ii)~For $m_\phi/m_\psi\gtrsim0.1$, the allowed bands raise with the mass ratio. The muon $g-2$ is more sensitive to the diagram with $\psi$ interacting with a photon [as depited in Fig.~\ref{fig:penguin&box}(b)] than to the other diagram, hence, the constraint on $- Q_{\psi}{\rm  Re}(g^*_{\mu R} g_{\mu L})/m_\psi$ is severer than the one on $+ Q_{\phi}{\rm  Re}(g^*_{\mu R} g_{\mu L})/m_\psi$.
Indeed in the large $m_\phi/m_\psi$ region, the suppressions from a large $\phi$ mass should be larger in diagrams with $\phi$ interacting with a photon and, hence, they require larger couplings to compensate the effect. 
We see that the mass ratio $m_\phi/m_\psi$ can go up to 200 (1000) along the $\phi$ ($\psi$) band, which corresponds to allowing $m_\phi$ to be as large as 20 (100)~TeV in the extreme situation.

\begin{figure}[tb]
\centering
\subfigure[]{
  \includegraphics[width=7.4cm]{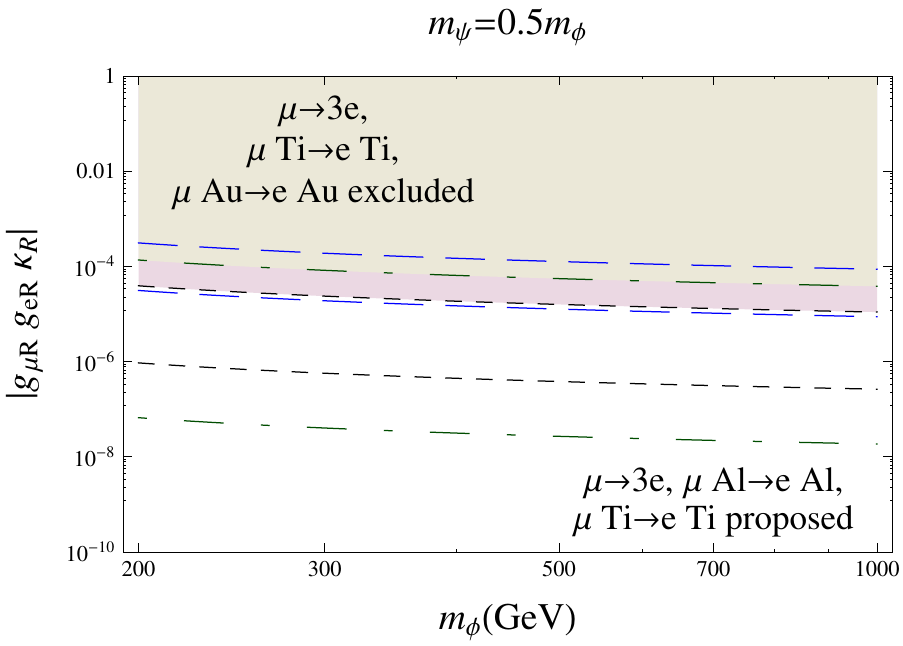}
}
\hspace{0.5cm}
\subfigure[]{
  \includegraphics[width=7.4cm]{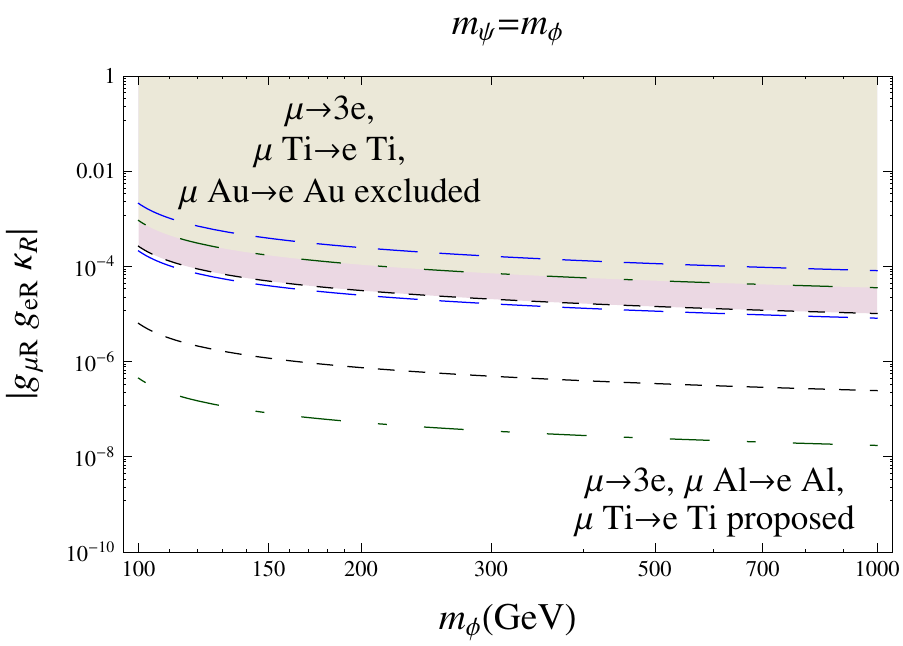}
}
\subfigure[]{
  \includegraphics[width=7.4cm]{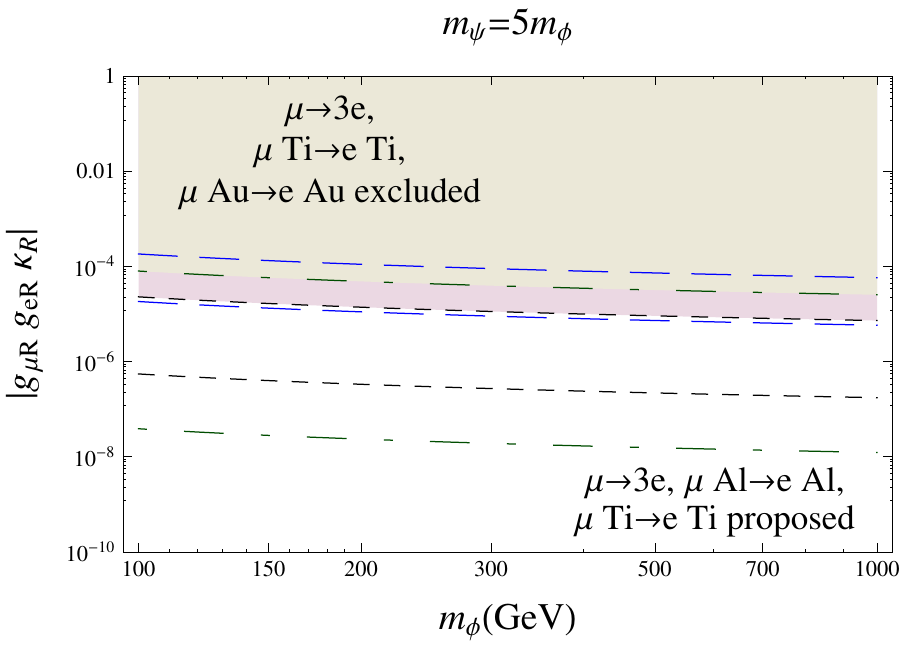}
}
\hspace{0.5cm}
\subfigure[]{
  \includegraphics[width=7.4cm]{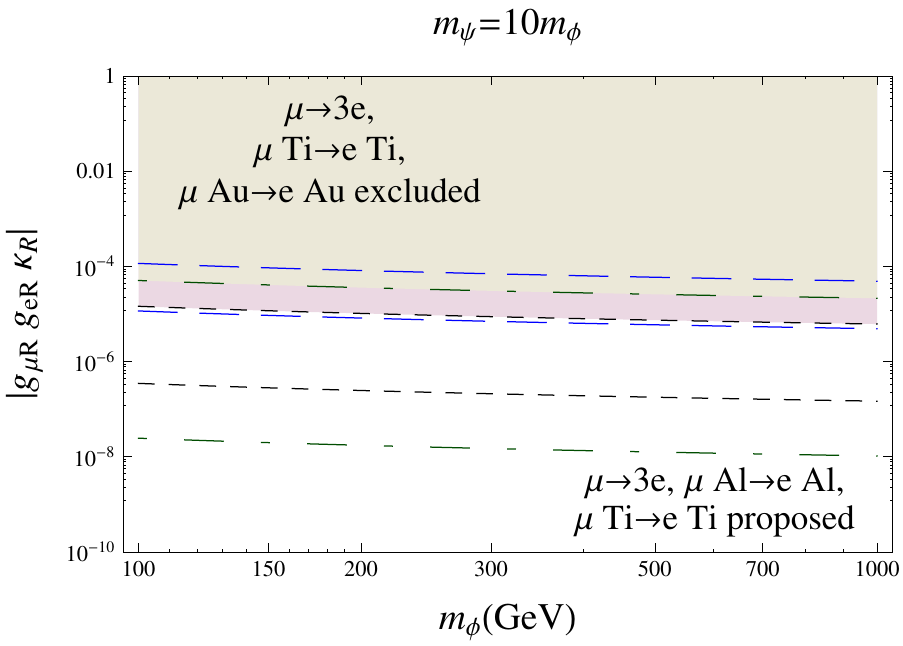}
}
\caption{(a)-(d): Parameter space excluded (projected) by various bounds (expecting sensitivities) on $\mu$ LFV processes through contributions from $Z$ penguins with different choices of mass ratios. Note that dashed, dot-dashed and short-dashed lines denote results from
$\mu\to 3 e$, $\mu {\rm Ti}\to e{\rm Ti}$ and $\mu {\rm Au(Al)}\to e {\rm Au(Al)}$ processes, respectively.
Note that these plots also apply to the $R\leftrightarrow L$ cases.} 
\label{fig:LFVpenguinZ}
\end{figure}

\begin{figure}[tb]
\centering
\subfigure[]{
  \includegraphics[width=7.5cm]{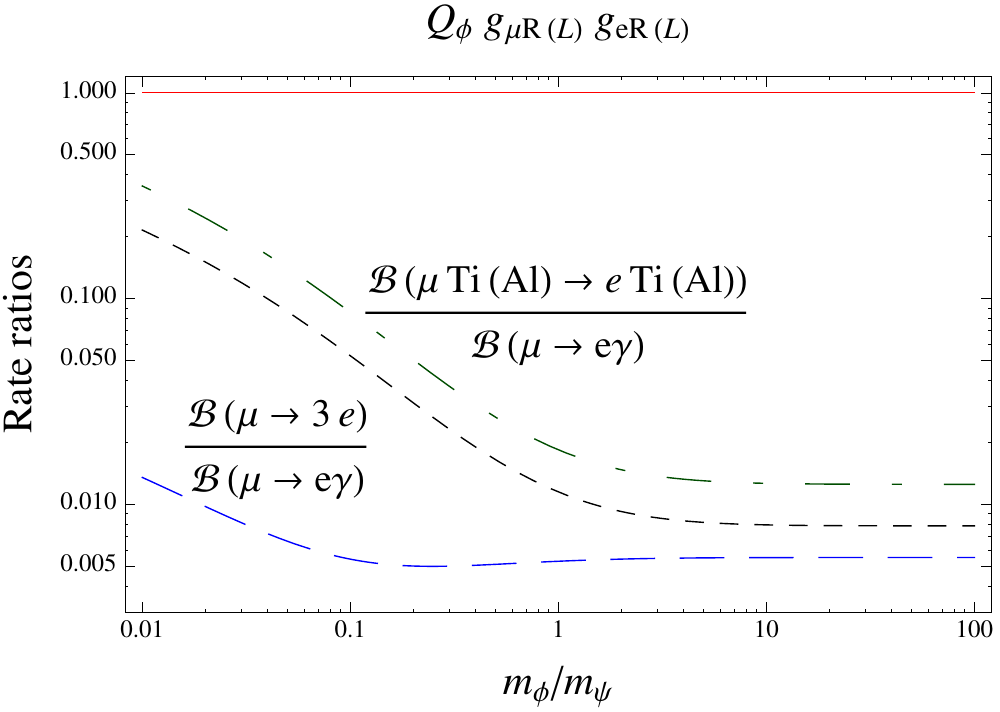}
}
\hspace{0.5cm}
\subfigure[]{
  \includegraphics[width=7.5cm]{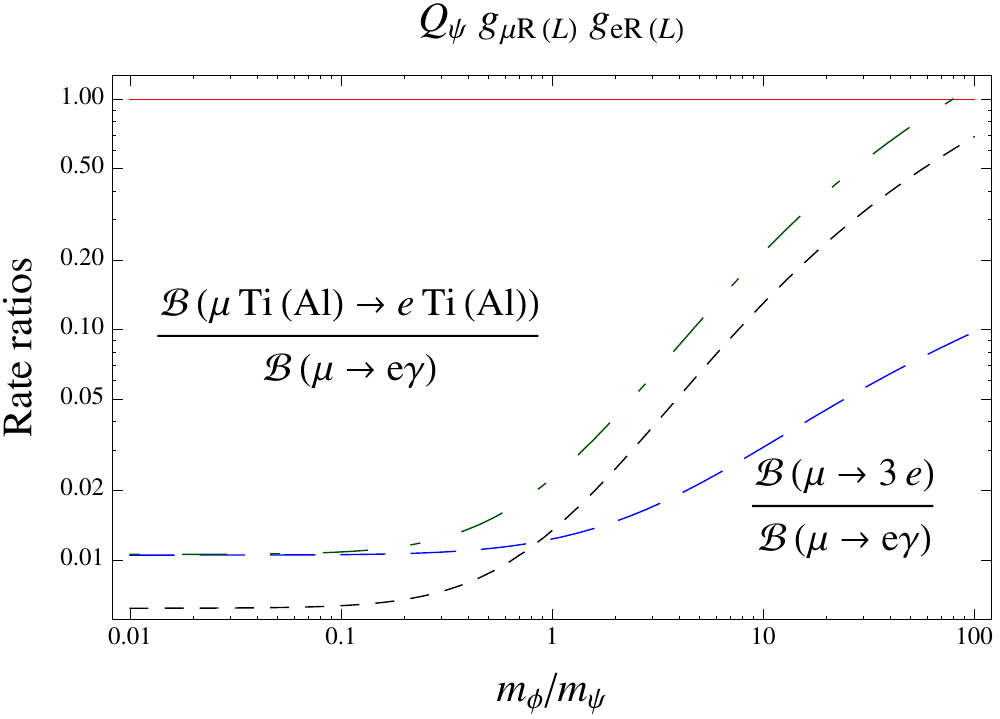}
}
\subfigure[]{
  \includegraphics[width=7.5cm]{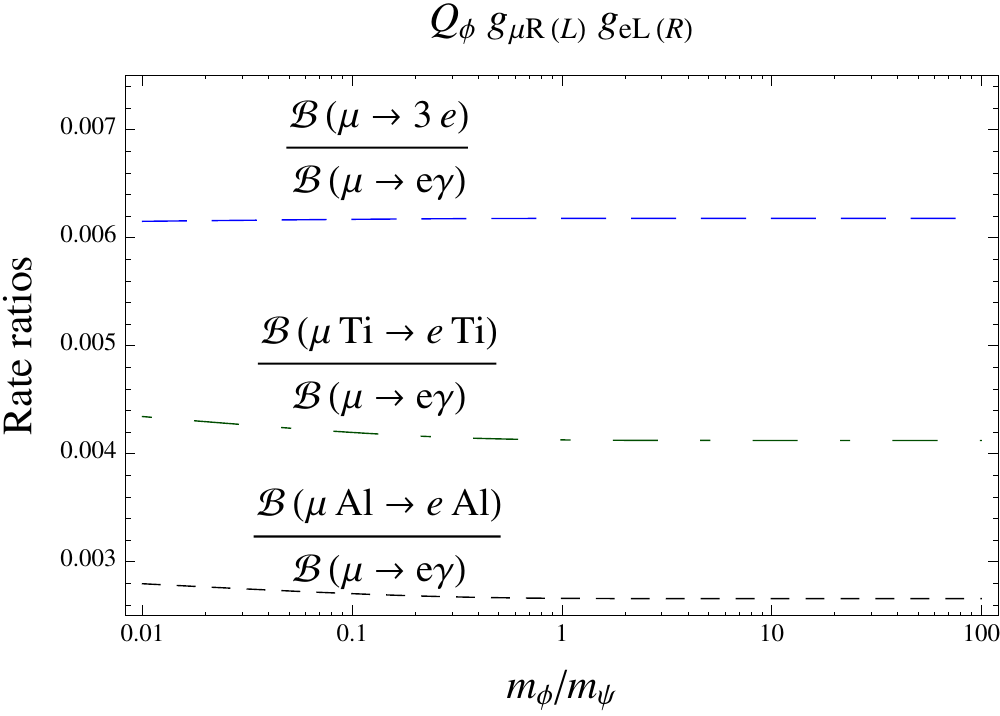}
}
\hspace{0.5cm}
\subfigure[]{
  \includegraphics[width=7.5cm]{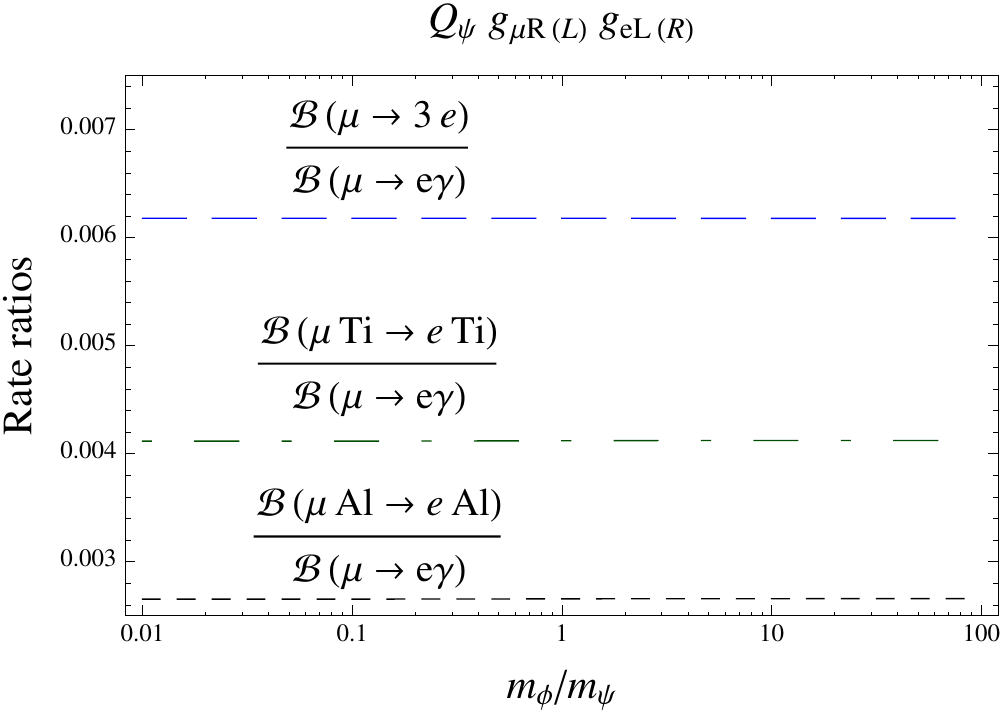}
}
\caption{Ratios of rates contributed from photonic penguins.}
\label{fig:LFVpenguinratio}
\end{figure}

As noted previously, in Fig.~\ref{fig:muon g-2 1}(b) we also show the excluded region of $|Q_{\phi,\psi} {\rm Im}(g^*_{\mu R} g_{\mu L})|/m_\psi$ from the muon EDM bound. We see that the bound is three order of magnitude higher that the allowed $\pm Q_{\phi,\psi}{\rm  Re}(g^*_{\mu R} g_{\mu L})/m_\psi$ bands. To constrain the former to the level of the latter, the EDM sensitivity needs to be improved. 
In fact, some proposed EDM searches (see, for example~\cite{Semertzidis:1999kv}) are aim at a 4 to 5 order of magnitude improvement on the sensitivity and may be able to probe the imaginary part of $g^*_{\mu R} g_{\mu L}$ better than its real part.

We now turn to $\mu$ LFV processes, including $\mu\to e \gamma$, $\mu\to 3 e$, $\mu {\rm Ti}\to e{\rm Ti}$ and $\mu {\rm Au(Al)}\to e {\rm Au(Al)}$ transitions.
In Fig.~\ref{fig:LFVpenguin} and \ref{fig:LFVpenguinZ}, we show the parameter space excluded by various bounds and the one corresponding to projections from the expected sensitivities on these $\mu$ LFV processes, through contributions from photonic and $Z$ penguin diagrams. 
To be specific, for the proposed sensitivities, the conservative values of the future sensitivities quoted in Table 1 are used. 
Note that the photonic penguins contribute to $\mu\to 3 e$ and $\mu N\to e N$  
through the so-called $F_2$ penguin, which is similar to those contributing to $\Delta a_\mu$ and $\mu\to e\gamma$, and the $F_1$ penguin, while 
the $Z$-penguins only contribute to $\mu\to 3 e$ and $\mu N\to e N$ decays. 
Note that the $Z$-penguin amplitudes contribute through the $|g_{\mu R(L)} g_{e R(L)} \Delta T_{3\psi}|$ and $|g_{\mu R(L)} g_{e R(L)} \kappa_{R(L)}|$ parts.  
The former contribution is a function of the mass ratio $m_\phi/m_\psi$, while the latter one depends on both $\phi$ and $\psi$ masses. 
The resulting constraints are plotted in Fig.~\ref{fig:LFVpenguin}(e), (f) and \ref{fig:LFVpenguinZ}.   
It should be noted that Fig.~\ref{fig:LFVpenguin} and \ref{fig:LFVpenguinZ} can still be useful if the experimental bounds change. For example, if a bound is reduced by a factor of $k$, the new plot can be easily updated by reducing the present plot by a factor of $\sqrt k$.

Note that the combinations of couplings $|Q_{\phi,\psi} g_{\mu R(L)} g_{e R(L)}|$, $|Q_{\phi,\psi} g_{\mu R(L)} g_{e L(R)}|$ (from photonic penguin) and $|g_{\mu R(L)} g_{e R(L)} \Delta T_{3\psi}|$, $|g_{\mu R(L)} g_{e R(L)} \kappa_{L(R)}|$ (from $Z$ penguin) have different sensitivities on the experimental constraints and the sensitivities change with $m_{\psi,\phi}$. 
For heavier $m_{\phi,\psi}$, the contributions from the photonic penguins decrease and the $Z$-penguin contributions, where the non-decoupling effect is working, dominate. 
By comparing Fig.~\ref{fig:LFVpenguin}(a), (b) to Fig.~\ref{fig:LFVpenguin} (e), (f) and \ref{fig:LFVpenguinZ},~\footnote{Note that the plotted quantities in these figures have different powers of $m_\psi$.}
we find that (i) in the range of $m_\psi\lesssim {\cal O}(100)$~GeV the photonic penguin contributions dominate over the $Z$-penguin ones,  (ii) in the range of the ${\cal O}(100)$~GeV $\lesssim m_\psi\lesssim {\cal O}(100)$~TeV, the photonic penguin contributions from the $Q_{\phi,\psi}g_{\mu R(L)} g_{e L (R)}$ terms dominates over the $Z$ penguin contributions, which are, however, still larger than the photonic penguin contributions from the $Q_{\phi,\psi}g_{\mu R(L)} g_{e R(L)}$ part, and for (iii) $m_\psi\gtrsim {\cal O}(100)$~TeV the $Z$ penguin contributions dominate. 
The role which $Z$ penguin plays is emphasized in \cite{Hirsch:2012ax}.

Since NP contributions to $\Delta a_\mu$ and the $\mu^+\to e^+\gamma$ decay are from similar diagrams, it will be useful to compare them. Using Fig.~\ref{fig:muon g-2 1}(b), \ref{fig:LFVpenguin}(c) and \ref{fig:LFVpenguin}(d),  the present data on $\Delta a_\mu$ and ${\cal B}(\mu^+\to e^+\gamma)$ lead to 
\be
\frac{g_{\mu R(L)} g_{e L(R)}}{g_{\mu R} g_{\mu L}}
=\frac{g_{e L(R)}}{g_{\mu L(R)}} 
\leq 6.1\times 10^{-5}
\simeq \lambda^6,
\label{eq:boundcase1}
\en
where we define $\lambda\equiv0.2$.   
This ratio is much smaller than any known coupling ratio and mixing angle among the first and second generations. 
For example, 
the mass ratio $m_e/m_\mu\sim\lambda^{3\sim4}$, quark mixing in CKM matirx $V_{ud}=\sin\theta_c\sim\lambda$, neutrino mixing $\sin\theta_{\nu 12}\sim\sqrt\lambda$ are all larger than the estimated $g_{eL(R)}/g_{\mu L(R)}$ coupling ratio.
It seems that the present case is unnatural. 

\begin{figure}[t]
\centering
\subfigure[]{
  \includegraphics[width=7.5cm]{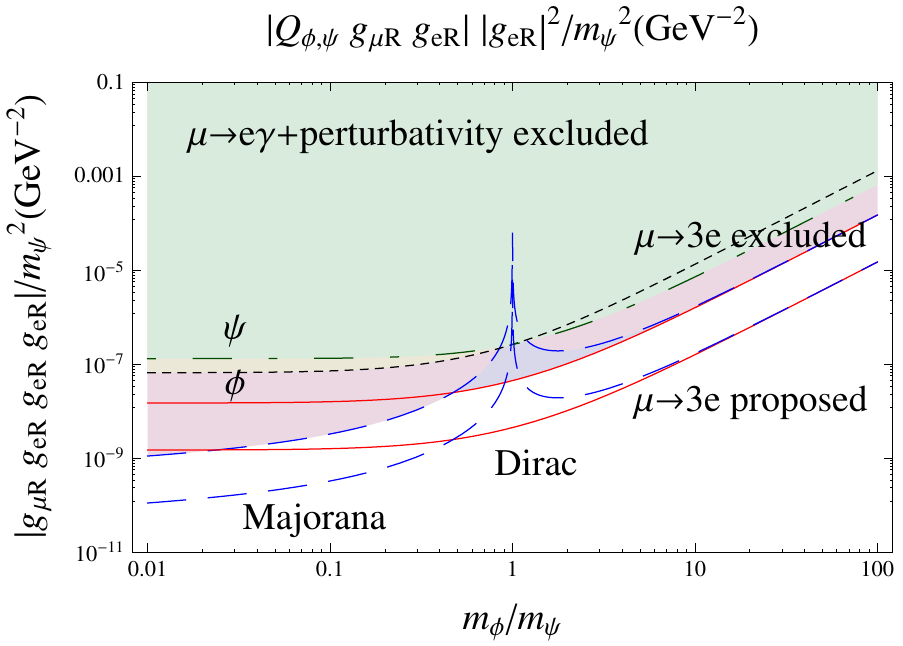}
}
\hspace{0.5cm}
\subfigure[]{
  \includegraphics[width=7.5cm]{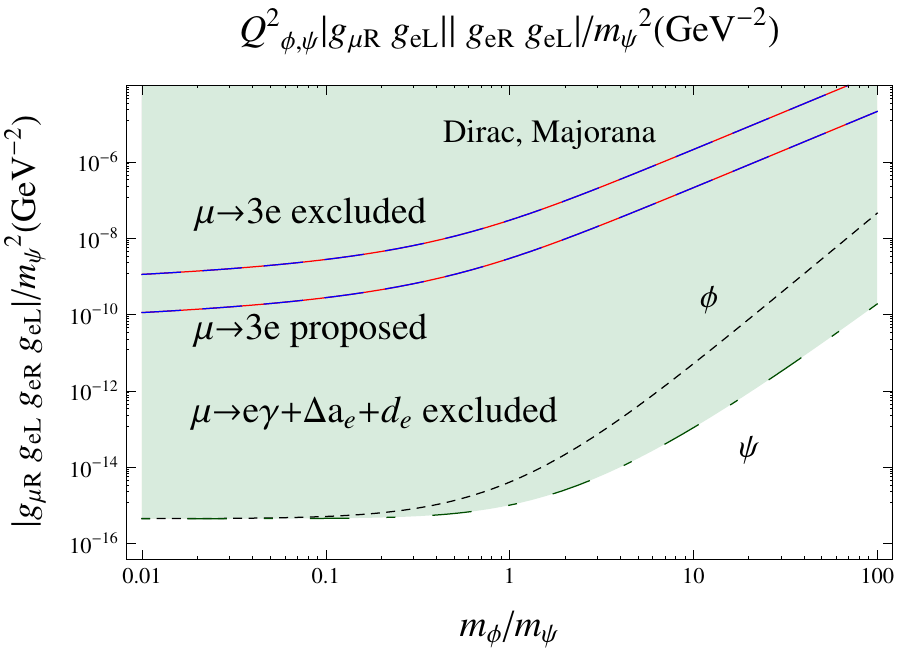}
}
\caption{Constraints on parameters which contribute through box diagrams to the $\mu^+\to 3 e$ process. Solid lines denote the $\mu \to 3e$ constraint or expectation in the Dirac case, while the dashed lines denote the Majorana case. Note that these plots also apply to the $R\leftrightarrow L$ cases.}
\label{fig:LFVbox}
\end{figure}

We see from Fig.~\ref{fig:LFVpenguin}(a)-(d) that the present bound from $\mu\to e\gamma$ surpasses all other bounds. 
In particular, even the parameter space to be probed by the proposed  $\mu\to 3 e$ sensitivity is mostly excluded  by the present $\mu\to e\gamma$ bound. 
This can be understood by using  
Fig.~\ref{fig:LFVpenguinratio}, where rate ratios of various modes through photonic penguins are given. 
We see that the ratios of  LFV rates with respect to the $\mu\to e\gamma$ rate are all less than unity. 
Furthermore, we recall that the present experimental bounds on LFV rates are of similar orders of magnitudes  
(see Table~1). 
Therefore, the present bound on the $\mu\to e\gamma$ rate provides the most severe constrain. 

Taking a closer look at Fig.~\ref{fig:LFVpenguinratio}, 
we see that, from Fig.~\ref{fig:LFVpenguinratio} (a) and (b), the $|Q_{\phi,\psi} g_{\mu R(L)} g_{e R(L)}|$ terms give ${\cal B}(\mu\to e \gamma)>{\cal B}(\mu N\to e N)\gtrsim{\cal B}(\mu \to 3 e)$ and, from Fig.~\ref{fig:LFVpenguinratio}~(c) and (d), the $|Q_{\phi,\psi} g_{\mu R(L)} g_{e L(R)}|$ terms give 
${\cal B}(\mu \to 3e)/{\cal B}(\mu\to e\gamma)\simeq 0.006$,
${\cal B}(\mu {\rm Ti}\to e {\rm Ti})/{\cal B}(\mu\to e\gamma)\simeq  0.004$ and
${\cal B}(\mu {\rm Al}\to e {\rm Al})/{\cal B}(\mu\to e\gamma)\simeq 0.003$, 
where the first ratio is consistent with Ref.~\cite{review}.
Hence, for $g_{\mu R(L)} g_{e L(R)}$ dominating models, the latest MEG bound implies
\be
&{\cal B}(\mu^+\to e^+e^+e^-)
\simeq 0.006\times {\cal B}(\mu^+\to e^+\gamma)\lesssim 1.4\times10^{-14},&
\non\\
&{\cal B}(\mu N\to eN)
\simeq{\cal O}(10^{-3})\times{\cal B}(\mu^+\to e^+\gamma)
\lesssim  {\cal O}(10^{-15}),&
\label{eq:bound1}
\en
for $N=$Au, Al and Ti.
The above expecting limits are about two to three orders of magnitudes below the present experimental sensitivities (see Table I) and make the searches on LFV in the muon sector challenging in this case.

As noted in Sec.~\ref{sec. Lint}, it is possible to have box diagrams with the electron line in Fig.~\ref{fig:penguin&box}(c), (d) replaced by a quark one and contribute to $\mu N\to e N$ conversions in some cases. 
The correlation to the $\mu \to 3e$ rate will be modified. 
We will discuss more on this situation in the discussion section.

Note that $Z$-penguins give different rate ratios (not shown in Fig.~\ref{fig:LFVpenguinratio}), with ${\cal B}(\mu {\rm Al}\to e {\rm Al})/{\cal B}(\mu\to 3e)\simeq 10$, ${\cal B}(\mu {\rm Ti}\to e {\rm Ti})/{\cal B}(\mu\to 3e)\simeq  20$ and ${\cal B}(\mu {\rm Au}\to e {\rm Au})/{\cal B}(\mu\to 3e)\simeq 40$ roughly independent of the masses $m_{\phi,\psi}$. 
This pattern is different from the photonic penguin case as shown Fig.~\ref{fig:LFVpenguinratio}.
These rate ratios 
will be useful for identifying the underlying NP contributions.

In Fig.~\ref{fig:LFVbox}, we show the constraints on parameters which contribute through box diagrams, as depicted in Fig.~\ref{fig:penguin&box} (c) and (d), to the $\mu^+\to 3 e$ process. 
Both Dirac and Majorana cases are shown. 
We see in Fig.~\ref{fig:LFVbox}~(a) that there is cancellation in the Majorana case and the sensitivity on the parameters is relaxed. 

Note that constraints on the same combinations of parameters can be obtained from penguin processes, including $\mu\to e\gamma$, $\Delta a_e$, EDM, and the purturbative bounds, as well. 
They are also shown in Fig.~\ref{fig:LFVbox}.   
We see that these constraints are usually much stronger than the ones from the box diagrams, except for $|g_{\mu R(L)} g_{eR(L)} g_{e R(L)} g_{eR(L)}|/m_\psi^2$ in the low $m_\phi/m_\psi$ region.
In particular, the  $\mu\to e\gamma$, $\Delta a_e$ and the electron EDM constrain $(Q_{\phi,\psi}|g_{\mu R(L)} g_{eL(R)}|/m_\psi) (Q_{\phi,\psi}g_{e R(L)} g_{eL(R)}|/m_\psi)$  
much deeper than $|g_{\mu R(L)} g_{eR(L)} g_{e L(R)} g_{eL(R)}|/m_\psi^2$ from the box diagrams.
Hence, in general, these box diagrams do not play a major role in the $\mu^+\to 3 e$ decay.

\subsection{Case II}

\begin{figure}[t]
\centering
  \includegraphics[width=7.5cm]{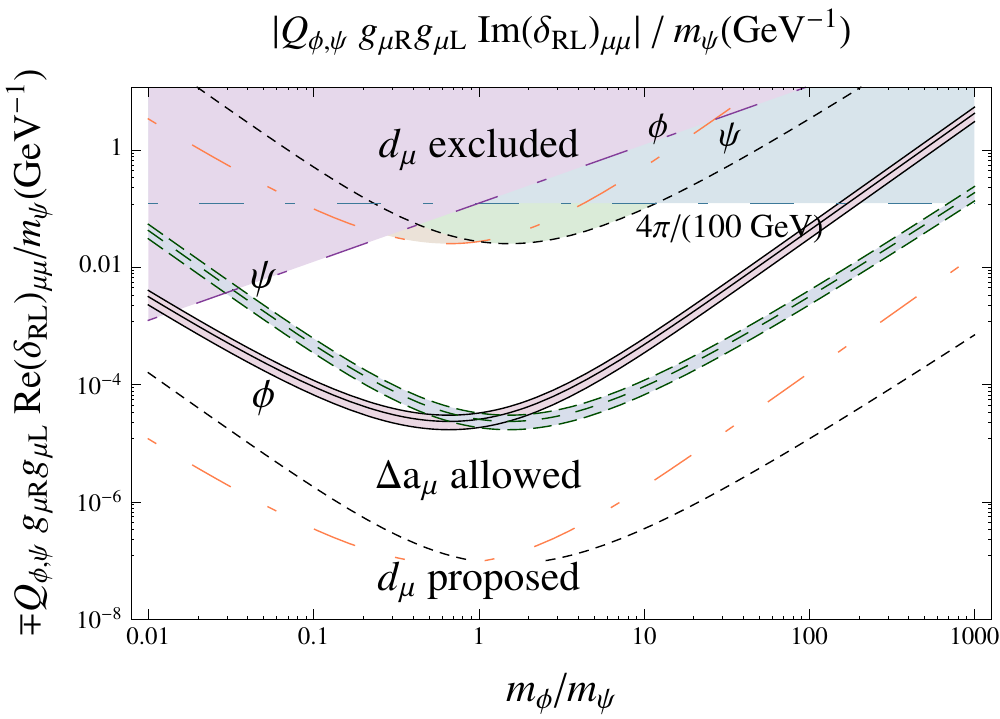}
\caption{
Allowed parameter space for
$\mp Q_{\phi,\psi}g_{\mu R} g_{\mu L}{\rm  Re}(\delta_{RL})_{\mu\mu}/m_\psi$ constrained by 
$\Delta a_\mu$ with exclusion of $|g_{\mu L} g_{\mu R}\delta_{RL}|>4\pi$ and $m_{\psi,\phi}>100$~GeV. Excluded parameter space (shaded regions with solid or dashed lines) of $|Q_{\phi,\psi}g_{\mu R} g_{\mu L}{\rm  Im}(\delta_{RL})_{\mu\mu}/m_\psi|$ from the muon EDM bound is also shown.}
\label{fig:muon g-2 2}
\end{figure}

We now turn to case II.
In Fig.~\ref{fig:muon g-2 2}, the allowed regions for
$\mp Q_{\phi,\psi}g_{\mu R} g_{\mu L}{\rm  Re}(\delta_{RL})_{\mu\mu}/m_\psi$ constrained by the measured
$\Delta a_\mu$, with exclusions of $|g_{\mu L} g_{\mu R}\delta_{RL}|>4\pi$ and $m_{\psi,\phi}<100$~GeV, are shown. Excluded and projected parameter space of $|Q_{\phi,\psi}g_{\mu R} g_{\mu L}{\rm  Im}(\delta_{RL})_{\mu\mu}|/m_\psi$ from the muon EDM bound and the expected sensitivity are also given on the same plot.
For the plots of the allowed regions for $\pm Q_{\phi,\psi} |g_{\mu L(R)}|^2/m_\psi^2$,  one is referred to Fig.~\ref{fig:muon g-2 1}(a), as they are common in both cases. .

\begin{figure}[tb]
\centering
\subfigure[]{
  \includegraphics[width=7.5cm]{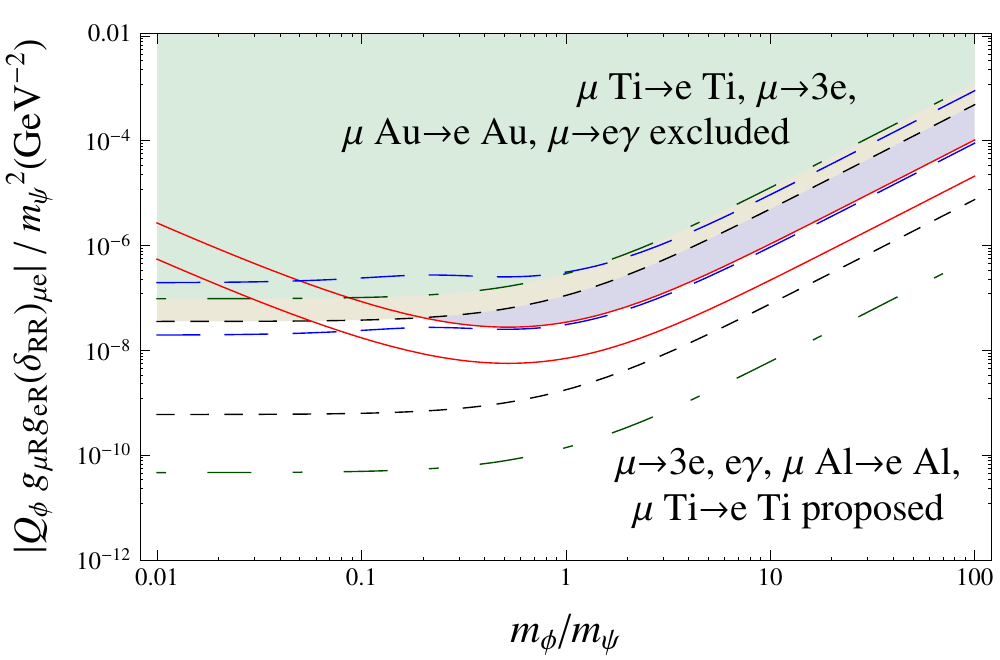}
}
\hspace{0.5cm}
\subfigure[]{
 \includegraphics[width=7.5cm]{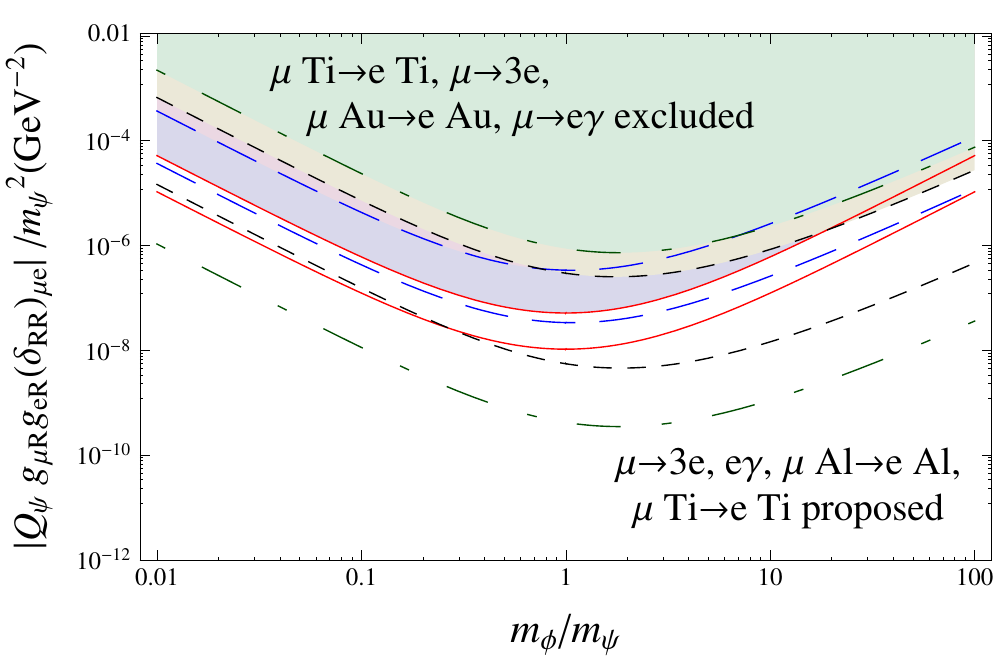}
}
\subfigure[]{
  \includegraphics[width=7.5cm]{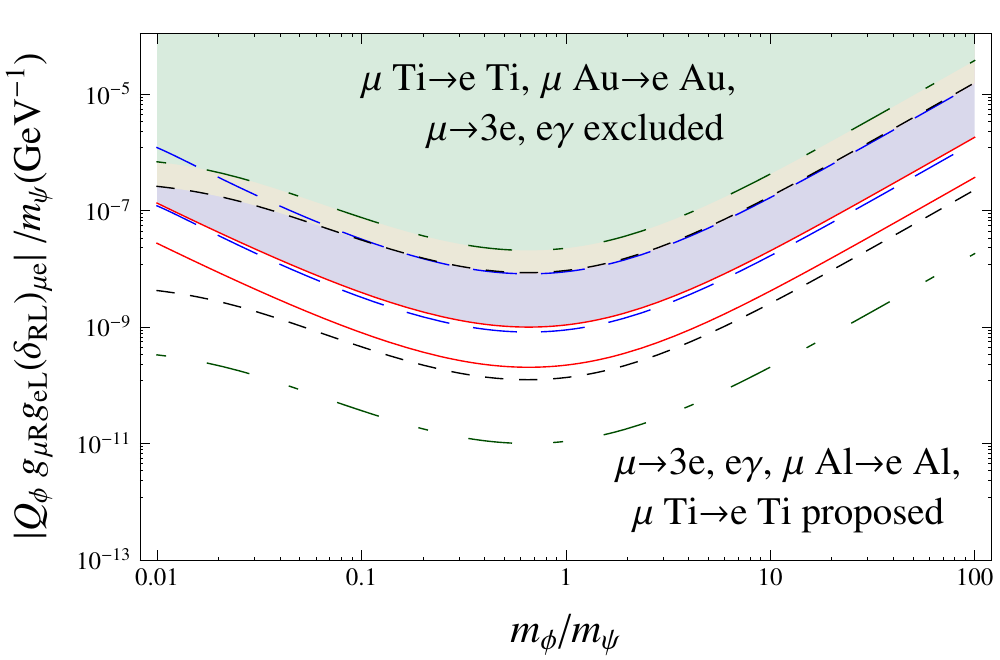}
}
\hspace{0.5cm}
\subfigure[]{
 \includegraphics[width=7.5cm]{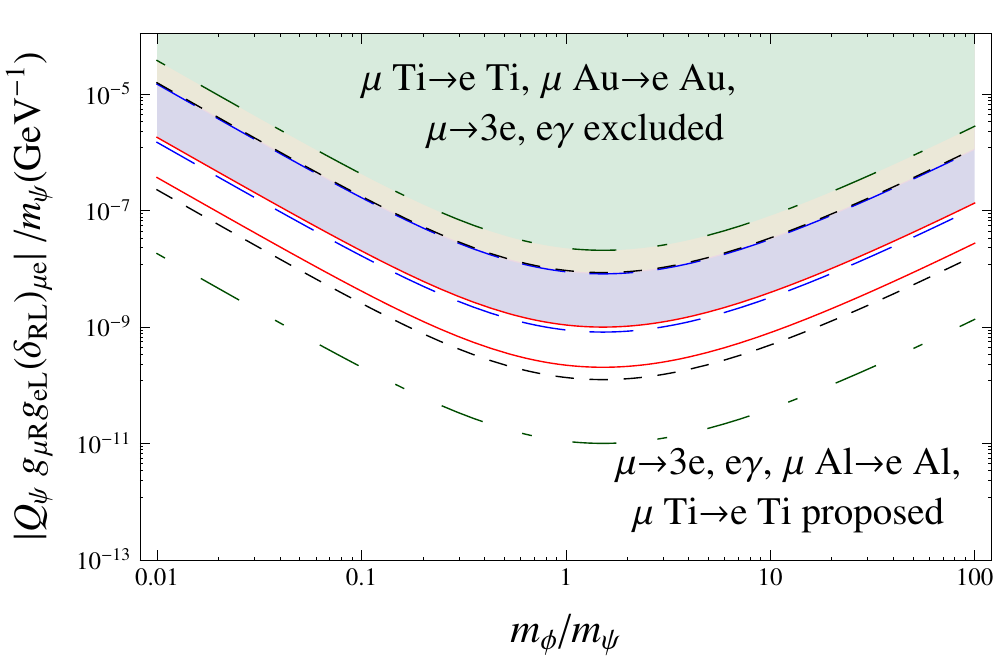}
}
\subfigure[]{
  \includegraphics[width=7.5cm]{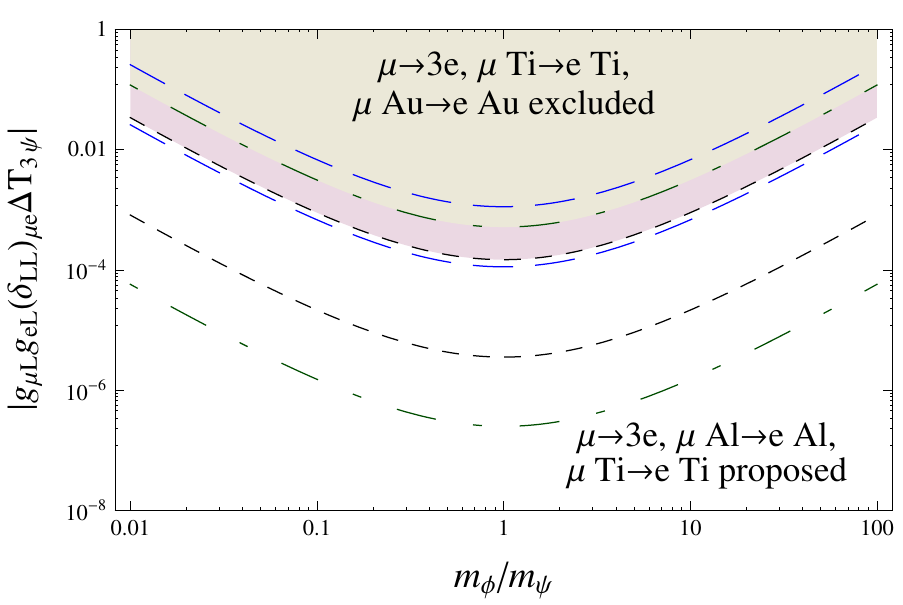}
}
\hspace{0.5cm}
\subfigure[]{
  \includegraphics[width=7.5cm]{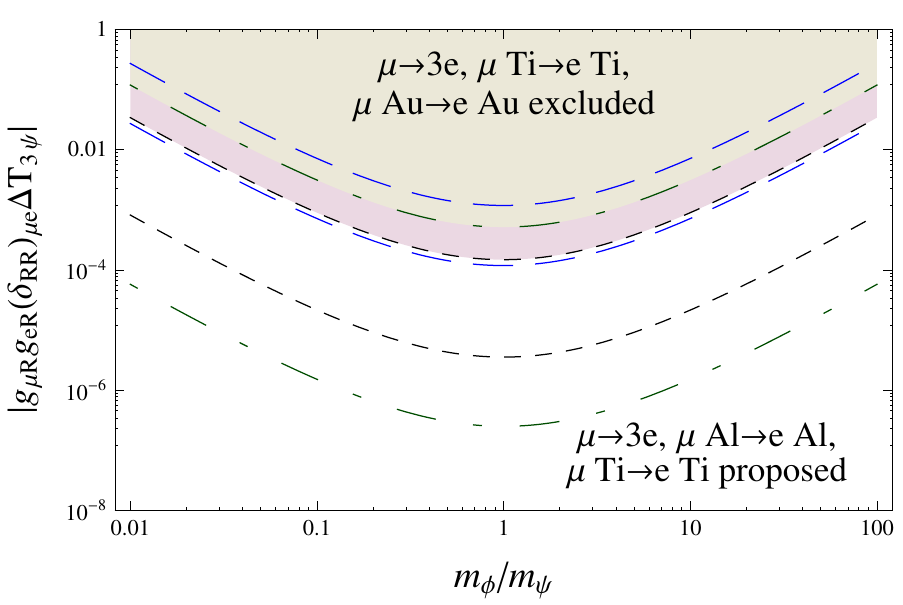}
}
\caption{Same as Fig.~\ref{fig:LFVpenguin}, but now in case II.
}
\label{fig:LFVpenguin2}
\end{figure}

Comparing Fig.~\ref{fig:muon g-2 2} to Fig.~\ref{fig:muon g-2 1}(b), we see that the allowed parameters in the upper $m_\phi/m_\psi$ region are similar. In contrast, they are relaxed substantially in the lower  $m_\phi/m_\psi$ region in the present case.  
To reproduce the measured $\Delta a_\mu$, we need to have ${\cal O}(10^{-2})\lesssim m_\phi/m_\psi\lesssim {\cal O}(10^{2,3})$, where the minimum of the mass ratio is much higher than the one in the previous case.
Recall that in case I, for $m_\phi/m_\psi\lesssim0.1$, the allowed parameter region for $\pm Q_{\phi,\psi}{\rm  Re}(g^*_{\mu R} g_{\mu L})/m_\psi$ are horizontal bands around $4\times 10^{-6}$ GeV$^{-1}$ and $m_\phi/m_\psi$ can be as low as ${\cal O}(10^{-5})$. From 
Fig.~\ref{fig:muon g-2 2}, we see that as we move downward along the $m_\phi/m_\psi$ axis, the bands for the allowed regions for $\mp Q_{\phi,\psi}g_{\mu R} g_{\mu L}{\rm  Re}(\delta_{RL})_{\mu\mu}/m_\psi$ bend upward in the low mass ratio region ($m_\phi/m_\psi< 1$) and the above parameters can be as large as $10^{-3}$~GeV$^{-1}$, which is three orders of magnitude higher than those in case I. We also note that the mass ratio $m_\phi/m_\psi$ cannot be smaller than $10^{-2}$ as the bands quickly run into the shaded rigions, which correspond to the excluded $m_\phi<100$~GeV and $|g_{\mu R} g_{\mu L}{\rm  Re}(\delta_{RL})_{\mu\mu}|>4\pi$ regions. 
In the present case, the mass of $\psi$ cannot be larger than a few tens TeV, 
while in case I it can be as high as few thousand TeV in the extreme situation. 
The built-in cancellation mechanism reduces the amplitudes effectively and a too heavy $\psi$ is incapable to produce a large enough $\Delta a_\mu$.
The effect of the cancellation is important in the low $m_\phi/m_\psi$ region and, consequently, relaxes the constraints on parameters.  In fact, we expect to see the very feature in other penguin contributing channels as well.

\begin{figure}[t]
\centering
\subfigure[]{
  \includegraphics[width=7.5cm]{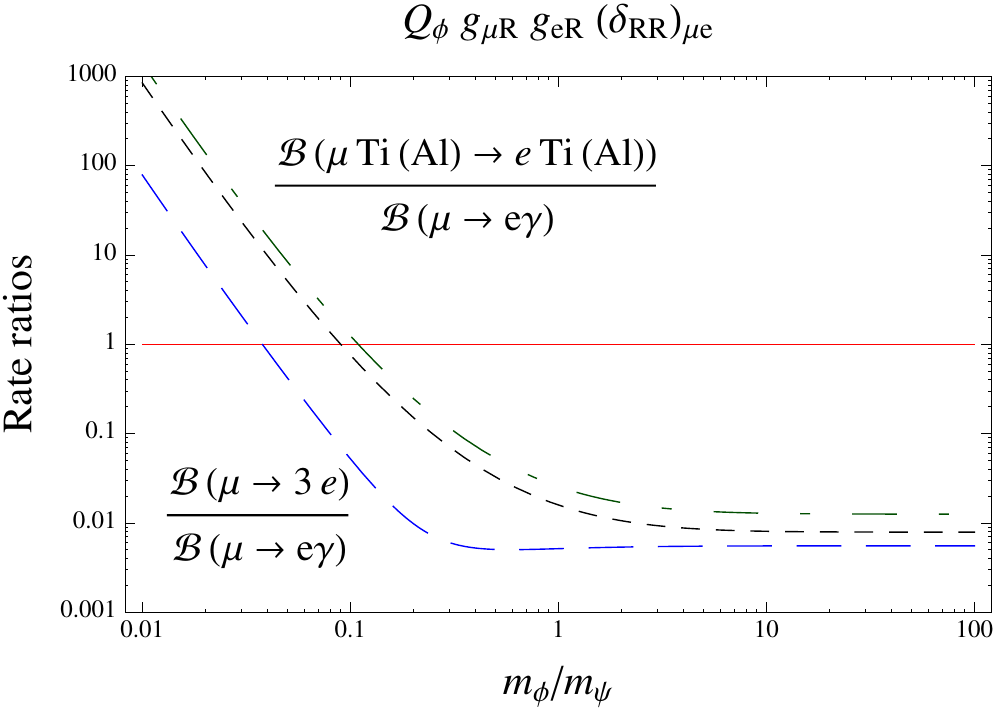}
}
\hspace{0.5cm}
\subfigure[]{
 \includegraphics[width=7.5cm]{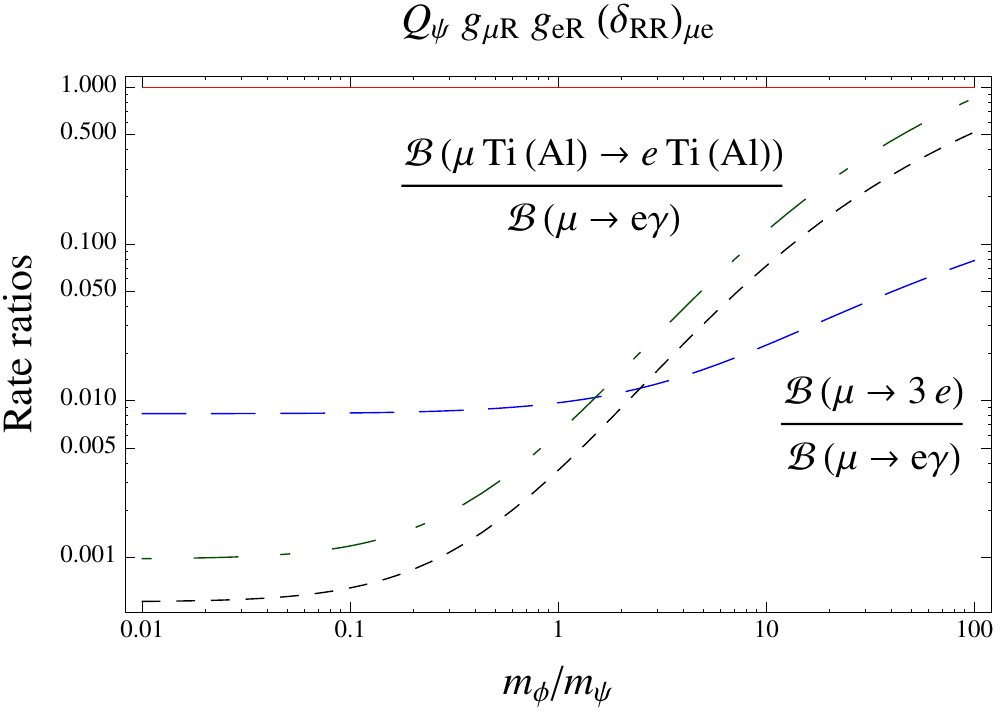}
}
\subfigure[]{
  \includegraphics[width=7.5cm]{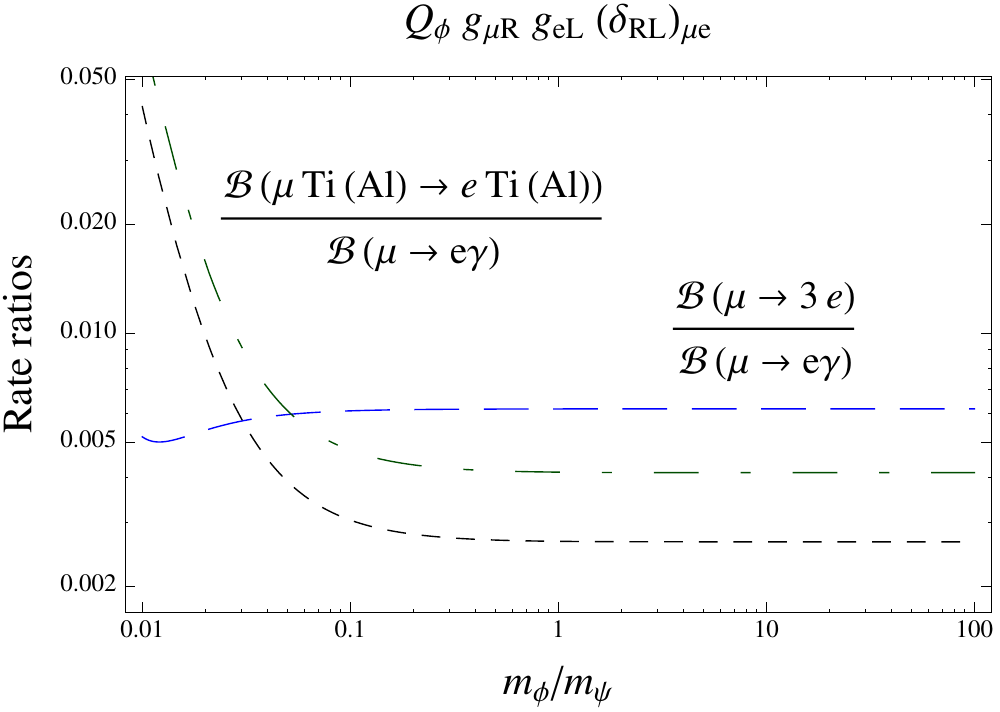}
}
\hspace{0.5cm}
\subfigure[]{
  \includegraphics[width=7.5cm]{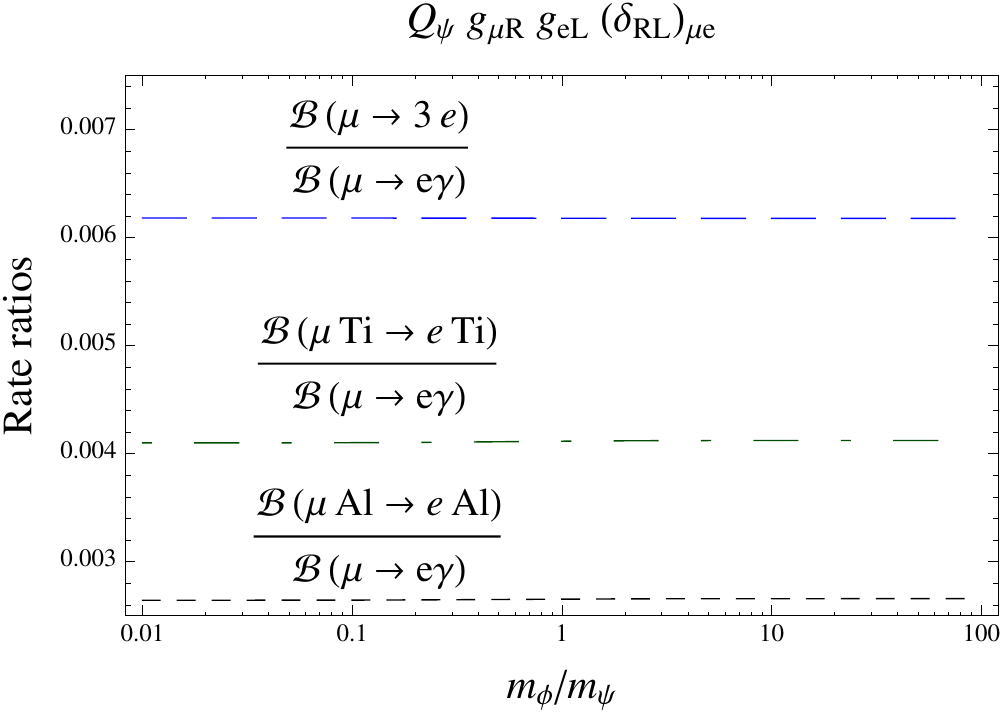}
}
\caption{Same as Fig.~\ref{fig:LFVpenguinratio}, but in case II.}
\label{fig:LFVpenguinratio2}
\end{figure}

In Fig.~\ref{fig:LFVpenguin2}, we show the constrained and projected parameter space through penguin contributions by considering the experimental bounds and the proposed sensitivities.~\footnote{For $Z$ penguin contributions, only those from $\Delta T_{3\psi}$ are shown, since the $\kappa_{L(R)}$ ones are highly suppressed.}  
We note that  
the photonic penguin contributions via the $\delta_{RL}$ term dominate over other contributions for $m_\psi$ below ${\cal O}(10^3)$ TeV. For $m_\psi$ beyond that the $Z$-penguin contribution takes over. However, from the previous discussion on the muon anomalous magnetic moment, we see that to account for the measured $\Delta a_\mu$, $m_\psi$ cannot be heavier than few tens TeV. Hence, the $Z$-penguin contribution will be subdominant in this case.

\begin{figure}[t]
\centering
\subfigure[]{
  \includegraphics[width=7.5cm]{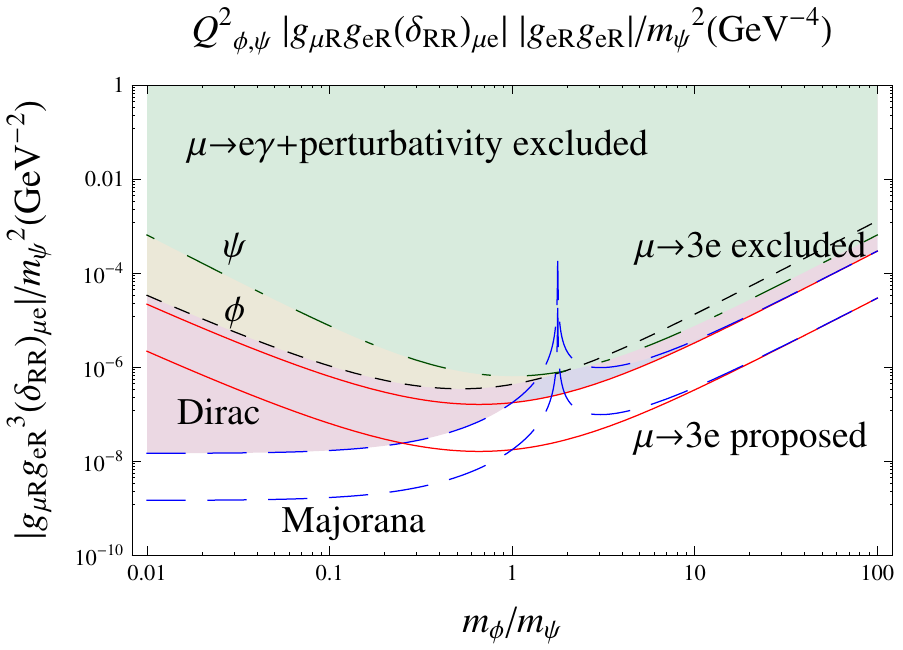}
}
\hspace{0.5cm}
\subfigure[]{
  \includegraphics[width=7.5cm]{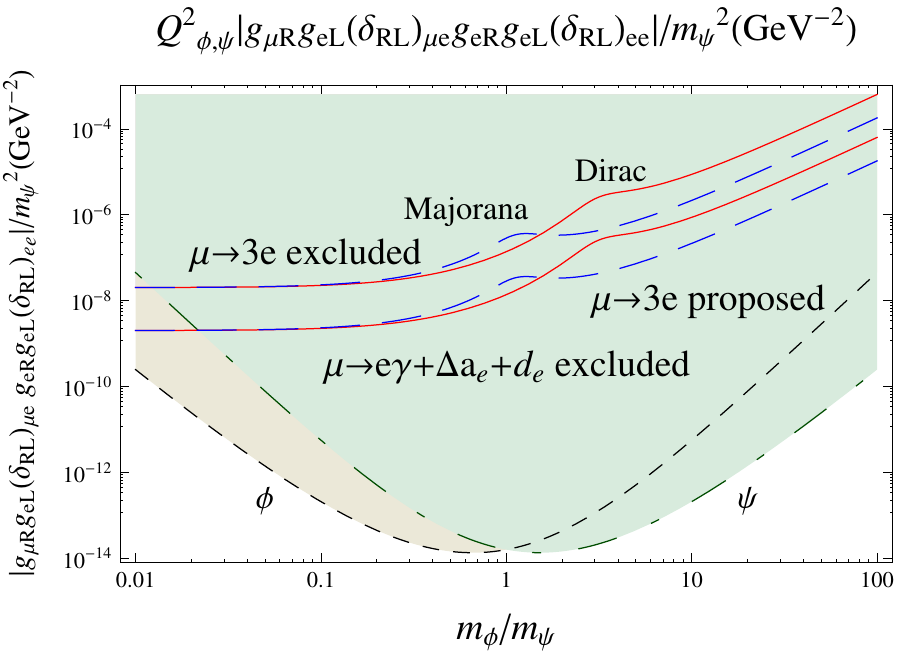}
}
\caption{Same as Fig.~\ref{fig:LFVbox}, but in case II.}
\label{fig:LFVbox2}
\end{figure}

By comparing the constraints from $\Delta a_\mu$ and ${\cal B}(\mu\to e\gamma)$ as
shown in Fig.~\ref{fig:muon g-2 2}, \ref{fig:LFVpenguin2}(c) and \ref{fig:LFVpenguin2}(d), 
we obtain
\be
\frac{g_{\mu R(L)} g_{e L(R)}{\rm Re}[(\delta_{R L(LR)})_{\mu e}]}{g_{\mu R} g_{\mu L}{\rm Re}[(\delta_{R L})_{\mu\mu}]}
=\frac{g_{e L(R)}}{g_{\mu L(R)}} \frac{{\rm Re}[(\delta_{R L(LR)})_{\mu e}]}{ {\rm Re}[(\delta_{R L})_{\mu\mu}]}
\leq 4.2\times 10^{-5}
\simeq \lambda^6.
\label{eq:boundcase2}
\en
If we estimate ${g_{e L(R)}}/{g_{\mu L(R)}}$ by using the lepton mass ratio $m_e/m_\mu\sim\lambda^{3\sim4}$, we see that 
a mixing angle ratio of ${{\rm Re}[(\delta_{R L(LR)})_{\mu e}]}/{ {\rm Re}[(\delta_{R L})_{\mu\mu}]}\lesssim \lambda^{2\sim3}$, 
which is not unnatural, can easily satisfy the above bound. In this respect, case II is more reasonable and natural than case I, 
where the coupling ratio is highly hierarchical [see, Eq.~(\ref{eq:boundcase1})].

It is interesting to see from Fig.~\ref{fig:LFVpenguin2} that the $\mu\to e\gamma$ bound is not always the most stringent one.
The bounds from $\mu\to 3 e$ and $\mu N\to e N$ in Fig.~\ref{fig:LFVpenguin2}(a) are almost the same as those in case I [see Fig.~\ref{fig:LFVpenguin}(a)], but the bound from $\mu\to e\gamma$ is relaxed up to more than two orders of magnitudes in the low $m_\phi/m_\psi$ region and becomes less severe than other bounds.
Similarly, comparing Fig.~\ref{fig:LFVpenguin2}(c) with Fig.~\ref{fig:LFVpenguin}(c), we see that in the low $m_\phi/m_\psi$ region both bounds from $\mu\to e\gamma$ and $\mu\to 3 e$ are relaxed up to three orders of magnitudes, while the changes on those from $\mu N\to eN$ are mild.
We can infer that, similar to the $\Delta a_\mu$ case, the $F_2$ (photonic) penguin amplitudes exhibit cancellations in amplitudes in the low $m_\phi/m_\psi$ region and relax the constraints from $\mu\to e\gamma$ significantly, while the cancellations in the $F_1$ penguin contributions in $\mu\to3 e$ and $\mu N\to e N$ processes are mild. 
As a result the bounds from $\mu N\to eN$ approach the $\mu\to e\gamma$ bound in this case, while in the previous case these two bounds are always apart.

The ratios of photonic penguin contributing rates plotted in Fig.~\ref{fig:LFVpenguinratio2}, show that ${\cal B}(\mu N \to e N)/{\cal B}(\mu\to e\gamma)$ and ${\cal B}(\mu \to 3e )/{\cal B}(\mu\to e\gamma)$ are enhanced compared with those in Fig.~\ref{fig:LFVpenguinratio}. In Fig.~\ref{fig:LFVpenguinratio2}(a) we see that the ratios can be enhanced up to three orders of magnitudes, in Fig.~\ref{fig:LFVpenguinratio2}(c) the ${\cal B}(\mu N \to e N)/{\cal B}(\mu\to e\gamma)$ ratio is enhanced by one order of magnitude, while the ${\cal B}(\mu \to 3e )/{\cal B}(\mu\to e\gamma)$ ratio does not change much. 
It is very interesting that the rate ratio ${\cal B}(\mu N \to e N)/{\cal B}(\mu\to e\gamma)$ from the $g_{\mu R(L)} g_{e L(R)}$ term is enhanced and different from case I.

We see in Fig.~\ref{fig:LFVpenguin2} that parameters with $\delta_{RL}$ [as shown in (c) and (d)] are most constrained by data. It is likely that these parameters give dominate contributions to LFV processes.
Using Fig.~\ref{fig:LFVpenguinratio2}(c) and (d) we find that the present bound on $\mu\to e\gamma$ allows 
\be
{\cal B}(\mu N\to e N)\lesssim  10^{-13}, 
\en
which is close to the present bounds (see Table~\ref{tab:expt bounds}). Therefore, the search on these processes could be very interesting.

In Fig.~\ref{fig:LFVbox2}, we show the constraints on parameters which contribute through box diagrams to the $\mu^+\to 3 e$ process in this case. Although we also see some relaxations on parameters,
the main conclusion remains similar to that in case I.

\section{Discussions}

\subsection{Flavor violating $Z$-decays}

Lepton flavor violating $Z\to \mu^\mp e^\pm$ decays are highly related to $\mu\to 3 e$, $\mu\to e\gamma$ and $\mu N\to e N$ processes via the $Z$ penguin contributions.
The $Z\to l'\bar l$ decay rate is given by
\be
\Gamma(Z\to l'\bar l)=\frac{m^5_Z}{24\pi}(|g^Z_{L'L}|^2+|g^Z_{R'R}|^2),
\label{eq: Z decays}
\en
where the dimensionful coefficient $g^Z_{M'M}$ is the same one used in Eq.~(\ref{eq:effective}).
Using Eq.~(\ref{eq: Z decays}) and the results in the previous section, we find that the present bounds from $\mu\to 3 e$, $\mu {\rm Ti}\to e{\rm Ti}$ and $\mu {\rm Au}\to e{\rm Au}$ processes constrain 
\be
{\cal B}(Z\to \mu^\mp e^\pm)\leq 4\times 10^{-13}, 7\times 10^{-14}, 6\times 10^{-15}, 
\en
respectively. Note that the above equation holds in both case I and II.
In any case, these constraints are far below the present limit, ${\cal B}(Z\to \mu^\mp e^\pm)\leq 1.7\times 10^{-6}$.

\subsection{Box diagrams involving quarks}

Form the explicit assignment of gauge quantum numbers of $\psi$ and $\phi$ as shown in Appendix~\ref{app: QN}, we see that it is possible to have $\phi$ couples to quarks [see Eq.~(\ref{eq: qqphi})]. These interaction can generate additional contributions to $\mu N\to e N$ conversion precesses through box diagrams similar to those in the $\mu \to 3 e$ ones as shown in Fig. 1(c) and (d), but with the (lower) electron line replaced by a quark line.

To have interaction with quarks, only rather specific choices of  $\psi$ and $\phi$ gauge quantum numbers are allowed (see Appendix~\ref{app: QN}). 
In particular, the case of $\psi_R: (1,1,1)$ and  $\phi_L: (1,2,-1/2)$ are of interest, for the fermion field is a SM singlet. 
In below we will use this case to illustrate the contributions from the additional box diagrams. 

The interacting lagrangian in this case is
\be
{\cal L}_{\rm int}=g_{lL}\bar \psi_R L_{L_i}\phi^*_{Li}+g_u \bar Q_{L i} u_R\phi_{L_i}+g_d\bar Q_{L i} d_R\epsilon_{ij}\phi^*_{L j}+h.c.,
\en
where $Q_L$ and $L_L$ are the quark and lepton doublets, respectively. 
Note that only the lower components of $L_L$ and $\phi_L$ are relevant to this analysis.
The box diagrams give
\be
g_{RV}(q)=0,
\quad
g_{LV}(d)=0,
\en
and
\be
g_{LV}(u)=\frac{1}{16\pi^2}\bigg\{ 
                            -\frac{1}{8}G(m^2_{\psi}, 0, m^2_{\phi},m^2_{\phi})
                            (g_{e L}^* g_{\mu L})( g_u^* g_u+g_d^* g_d) \bigg\},                           
\en
where quark masses have been neglected. 
Note that the box diagrams also give the so-called $g_{LP}(u)$ term, which, however, does not contribute to conversion rates~\cite{KKO}.
The resulting $\mu N\to e N$ conversion rates can be calculated using Eq.~(\ref{eq: conv}). 

\begin{figure}[t]
\centering
  \includegraphics[width=7.5cm]{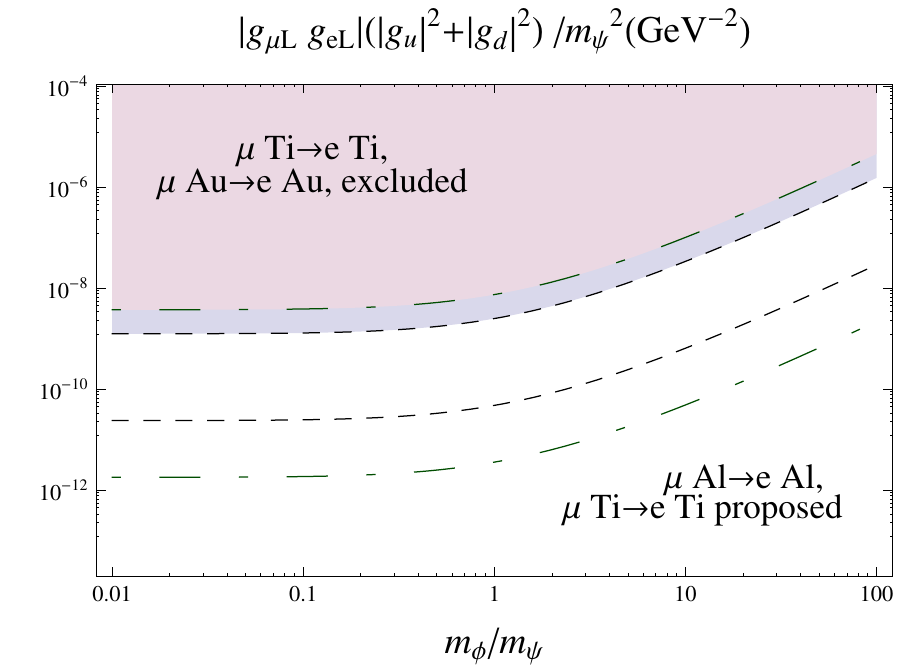}
\caption{Allowed parameter space for $|g_{\mu L} g_{eL}|^2(|g_u|^2+|g_d|^2)/m_\psi^2$ constrained by $\mu\to e$ conversion data.
Note that
dot-dashed and short dashed lines denote 
constraints from $\mu {\rm Ti}\to e{\rm Ti}$ and $\mu {\rm Au(Al)}\to e {\rm Au(Al)}$ conversion bounds, respectively.} 
\label{fig:boxphipsiquark}
\end{figure}

Experimental limits on conversion rates are used to constrain couplings and masses. The result is shown in Fig.~\ref{fig:boxphipsiquark}.
The correlation between $\mu N\to e N$ conversions and the $\mu^+\to 3e$ decay are lost.
In fact, we see that the constraints on $|g_{\mu L} g_{eL}|(|g_u|^2+|g_d|^2)/m_\psi^2$ from present limits on $\mu\to e$ conversion rates are similar to the constraints on $|g_{\mu L} g_{eL}| |g_{eL} g_{eL}|/m_\psi^2$ from the $\mu^+\to 3e$ bound (see Fig.~\ref{fig:LFVbox}).
Therefore, we may be able to see $\mu N\to e N$ conversions sooner than the $\mu^+\to 3e$ decay, if $g_{u,d}$ is larger than $g_{eL}$, and vice versa.
The $\mu N\to e N$ conversion rates need not be highly suppressed as noted in Sec.~\ref{sec: case I}.

\subsection{Some other cases}
A similar analysis can be preformed by replacing the spin-0 particle by a spin-1 one in the loops. 
It will be interesting to compare it to the present work.
However, gauge invariant and triplet vector couplings will complicate the analysis. 
The study will be given else where.

We expect to find results similar to case II, but with cancellation at work in low $m_\psi$ region, 
if we introduce the built-in cancellation mechanism in the $\psi$ sector, instead of in the $\phi$ sector.

\section{Conclusions}

In conclusion,
we use a model independent approach in this analysis, where these processes are considered to be loop-induced by exchanging spin-1/2 and spin-0 particles.
We explore two complementary cases, which has no or has an internal (built-in) cancellation mechanism in amplitudes.
Our main results are as follows:
\begin{itemize}

\item[(a)] Bounds from rates are used to constrain parameters, such as coupling constants and masses. These constraints can be easily updated by simple scalings, if the experimental situations change.

\item[(b)] The muon $g-2$ data favors non-chiral interactions.

\item[(c)] In $\mu^+\to e^+e^-e^-$ and $\mu^- N\to e^- N$ processes, the $Z$-penguin diagrams may play some role, while the box diagrams contributions to the $\mu^+\to e^+e^-e^-$ rate are usual highly constrained. 

\item[(d)] $Z$-penguin contributions can be constrained from $\mu^+\to e^+\gamma$ and $\mu^- N\to e^- N$ bounds. It can then be used to constrain the $Z\to e^\mp\mu^\pm$ rate by 7 to 8 orders of magnitudes lower than the present experimental bound.

\item[(d)] In the first case (without any built-in cancellation mechanism),  using the recent $\mu^+\to e^+\gamma$ bound, we find that $\mu^+\to e^+e^-e^-$ and $\mu^- N\to e^- N$ rates are bounded below the present experimental limits by two to three orders of magnitudes in general.  In some cases, the above expectation on low $\mu^- N\to e^- N$ rates can be relaxed, as additional box diagrams involving quarks contribute to $\mu^-N\to e^- N$ processes.

\item[(e)] Furthermore, by comparing $\Delta a_\mu$ and ${\cal B}(\mu\to e\gamma)$ data, the couplings of $g_\mu$ and $g_e$ are found to be highly hierarchical [see Eq.~(\ref{eq:boundcase1})]. 
Additional suppression mechanism should be called for.

\item[(f)] In the second case (with a built-in cancellation mechanism), mixing angles can provide additional suppression factors to satisfy the $\Delta a_\mu$ and ${\cal B}(\mu\to e\gamma)$ bounds without relay only on highly hierarchical $g_e$ and $g_\mu$ couplings.

\item[(g)]  In addition, although the $\mu^+\to e^+e^-e^-$ rate remains suppressed, 
the bounds on $\mu^- N\to e^- N$ rates, implicated from the MEG $\mu^+\to e^+\gamma$ bound, can be relaxed significantly in the second case and can be just below the present experimental limits.

\end{itemize}

\vskip 1.71cm {\bf Acknowledgments}

This research was supported in part by the National
Science Council of R.O.C. under grant No NSC97-2112-M-033-002-MY3 and NSC100-2112-M-033-001-MY3.

\appendix

\section{Gauge quantum numbers of $\phi$ and $\psi$}\label{app: QN}

The $\psi-\phi-l$ lagrangian,
\be
{\cal L}_{\rm int}= 
g'_L (\bar\psi_{R } \phi_{L}^*)_i (L_L)_i+g'_R \bar \psi_{L} \phi_{R }^* l_R
+h. c.,
\en
where $i$ is the weak isospin index, is gauge invariant under the SM gauge transformation. Recall that the lepton quantum numbers under SU(3)$\times$SU(2)$\times$U(1) are given by
\be
L_L:(1,2,-\frac{1}{2}),
\quad
l_R: (1,1,-1).
\en
The gauge invariant requirement implies that we must have the following quantum number assignments for these combinations:
\be
\bar\psi_R\phi_L^*: (1,2,\frac{1}{2}),
\quad
\bar\psi_L\phi_R^*: (1,1,1).
\en
Consequently, the gauge quantum numbers of $\psi$ and $\phi$ are related as following:
\be
\psi_R: (c_R, 2I_R+1, Y_R),
\quad
\phi_L: (\bar c_R, 2(I_R\pm \frac{1}{2})+1, Y_R-\frac{1}{2}),
\non\\
\psi_L: (c_L, 2I_L+1, Y_L),
\quad
\phi_R: (\bar c_L, 2I_L+1, Y_R-1).
\label{eq: QN}
\en
Some examples for the assignments of the quantum numbers of $\psi_{L,R}$ and $\phi_{L,R}$ are given in Table~\ref{tab:QN}.

Note that in the cases of $I_R=0$, $Y_R=0$ and $I_L=1/2$, $Y_L=1/2$, $\phi_L$ and $\phi_R$ can couple to quarks, respectively, through
\be
\bar Q_{L i} u_R \phi_{L i}, 
\quad
\bar Q_{L i} d_R \phi_{R i},
\quad
\bar Q_{L_i} u_R \epsilon_{ij}\phi^\dagger_{R j},
\quad
\bar Q_{L_i} d_R \epsilon_{ij}\phi^\dagger_{Lj},
\label{eq: qqphi}
\en
where $\epsilon_{ij}$ is the antisymmetric tensor. It is easy to see that the above terms are indeed gauge invariant by using 
$\bar Q_L q_R: (1,2,-1/6+Q_q)$ and Eq.~(\ref{eq: QN}).

\begin{table}[t]
\caption{Some examples for the assignment of the quantum numbers of $\psi_{L,R}$ and $\phi_{L,R}$.}
 \label{tab:QN}
\begin{ruledtabular}
\begin{tabular}{ c c | cc}
     $\psi_R$
    & $\phi_L$
    &$\psi_L$
    & $\phi_R$
    \\
    \hline
     $(1,1,Y_R)$
    & $(1,2,Y_R-\frac{1}{2})$
    &  $(1,1,Y_L)$
    & $(1,1,Y_L-1)$
     \\    
    $(1,2,Y_R)$
    &  $(1,1,Y_R-\frac{1}{2})$
    & $(1,2,Y_L)$
    & $(1,2,Y_L-1)$
    \\
    $(3 (\bar 3),1,Y_R)$
    & $(\bar 3 (3),2,Y_R-\frac{1}{2})$
    & $(3 (\bar 3),1,Y_L)$
    & $(\bar 3 (3),1,Y_L-1)$
    \\    
    $(3 (\bar 3),2,Y_R)$
    &  $(\bar 3 (3),1,Y_R-\frac{1}{2})$
    &  $(3 (\bar 3),2,Y_L)$
    &  $(\bar 3 (3),2,Y_L-1)$
    \\
\end{tabular}
\end{ruledtabular}
\end{table}


\section{Loop functions and input parameters}\label{app:FG}

The loop functions used in this work are defined as
\be
F_{1}(a,b)&=&\frac{1}{12(a-b)^4}\left(2 a^3+3 a^2 b-6 a b^2+b^3+6 a^2 b \ln\frac{b}{a}\right),
\non\\
F_{2}(a,b)&=&\frac{1}{2(a-b)^3}\left(-3 a^2+4 a b- b^2-2 a^2  \ln\frac{b}{a}\right),
\non\\
F_{3}(a,b)&=&\frac{1}{2(a-b)^3}\left(a^2- b^2+2 a b  \ln\frac{b}{a}\right),
\non\\
G_{1}(a,b)&=&\frac{1}{36(a-b)^4}\left(-(a-b)(11 a^2-7ab+2b^2)-6a^3 \ln\frac{b}{a}\right),
\non\\
G_{2}(a,b)&=&\frac{1}{36(a-b)^4}\left(-(a-b)(16 a^2-29ab+7b^2)-6a^2(2a-3b) \ln\frac{b}{a}\right),
\non\\
G_{3}(a,b)&=&\frac{1}{36(a-b)^5}\left(-(a-b)(17 a^2+8ab-b^2)-6a^2(a+3b) \ln\frac{b}{a}\right),
\non\\
F_Z(a_1,a_2,b,b,c)&=&-\frac{a_1(2\sqrt{a_1 a_2}-a_1)}{2(a_1-a_2)(a_1-b)} \ln \frac{a_1}{c}
                                       +\frac{a_2(2\sqrt{a_1 a_2}-a_2)}{2(a_1-a_2)(a_2-b)} \ln \frac{a_2}{c}
\non\\
                               &&  -\frac{b(2\sqrt{a_1 a_2}-b)}{2(a_1-b)(a_2-b)} \ln \frac{b}{c}
\non\\
F_Z(a,a,b_1,b_2,c)&=&-\frac{3}{4}
                                      +\frac{a^2}{2(a-b_1)(a-b_2)} \ln \frac{a}{c}
                                      -\frac{b_1^2}{2(a-b_1)(b_1-b_2)} \ln \frac{b_1}{c}
\non\\
                                && +\frac{b_2^2}{2(a-b_2)(b_1-b_2)} \ln \frac{b_2}{c},
\non\\
G_Z(a_1,a_2,b)&=&\frac{a_1\sqrt{a_1 a_2}}{(a_1-a_2)(a_1-b)} \ln \frac{a_1}{b}
                                     -\frac{a_2\sqrt{a_1 a_2}}{(a_1-a_2)(a_2-b)} \ln \frac{a_2}{b},
\non\\
F(a,b,c,d)&=&\frac{b\sqrt{a b}}{(a-b)(b-c)(b-d)}\ln \frac{b}{a}
                       -\frac{c\sqrt{a b}}{(a-c)(b-c)(c-d)}\ln \frac{c}{a}
\non\\
                 &&+\frac{d\sqrt{a b}}{(a-d)(b-d)(c-d)}\ln \frac{d}{a},
\non\\
G(a,b,c,d)&=&-\frac{b^2}{(a-b)(b-c)(b-d)}\ln \frac{b}{a}
                       +\frac{c^2}{(a-c)(b-c)(c-d)}\ln \frac{c}{a}
\non\\
                 && -\frac{d^2}{(a-d)(b-d)(c-d)}\ln \frac{d}{a}.                       
\en
Note that these loop functions are dimensionful and the dimension of $G_3$ is different from others.
We do not have the expression of $F_Z(a_1,a_2,b_1,b_2,c)$, since in Sec.~\ref{subsec:Z} only $a_1=a_2=a$ and/or $b_1=b_2=b$ are needed. 
Both expressions of $F_Z$ give identical result in the $a_1=a_2=a$ and $b_1=b_2=b$ case.

The numerical values of $D$, $V$ and $\omega_{\rm capt}$ used in Eq.~(\ref{eq: conv}) are collected in Table~\ref{tab:parameters}.

\begin{table}[t]
\caption{Parameters of overlap integrates and total capture rates $\omega_{\rm capt}$ taken from \cite{KKO,capt}.}
 \label{tab:parameters}
\begin{ruledtabular}
\begin{tabular}{ l c  c c c}
~~~~~~
    & $D(m_\mu^{5/2})$
    & $V^{(p)}(m_\mu^{5/2})$
    & $V^{(n)}(m_\mu^{5/2})$
    & $\omega_{\rm capt}(10^{6} s^{-1})$
    \\
    \hline
${}^{27}_{13}{\rm Al}$
    & 0.0362
    &  0.0161
    &  0.0173
    & 0.7054
    \\    
${}^{48}_{22}{\rm Ti}$
    & 0.0864
    &  0.0396
    &  0.0468
    & 2.59
    \\
${}^{197}_{79}{\rm Au}$
    & 0.189
    &  0.0974
    &  0.146
    & 13.07
    \\    
${}^{205}_{81}{\rm Tl}$
    & 0.161
    &  0.0834
    &  0.128
    & 13.90
    \\
\end{tabular}
\end{ruledtabular}
\end{table}


\end{document}